\pdfoutput=1
\RequirePackage{ifpdf}
\ifpdf 
\documentclass[pdftex]{sigma}
\else
\documentclass{sigma}
\fi

\numberwithin{equation}{section}

\begin{document}


\newcommand{\arXivNumber}{1411.7595}

\renewcommand{\PaperNumber}{028}

\FirstPageHeading

\ShortArticleName{From Principal Series to Finite-Dimensional Solutions of the Yang--Baxter Equation}

\ArticleName{From Principal Series to Finite-Dimensional\\ Solutions of the Yang--Baxter Equation}

\Author{Dmitry CHICHERIN~$^{\dag}$, Sergey E.~DERKACHOV~$^{\ddag}$ and Vyacheslav P.~SPIRIDONOV~$^\S$}

\AuthorNameForHeading{D.~Chicherin, S.E.~Derkachov and V.P.~Spiridonov}

\Address{$^\dag$~LAPTH, UMR 5108 du CNRS, associ\'ee \`a l'Universit\'e de Savoie,
Universit\'{e} de Savoie, CNRS,\\
\hphantom{$^\dag$}~B.P.~110, F-74941 Annecy-le-Vieux, France}
\EmailD{\href{mailto:chicherin@lapth.cnrs.fr}{chicherin@lapth.cnrs.fr}}

\Address{$^\ddag$~St.~Petersburg Department of Steklov Mathematical Institute
of Russian Academy of Sciences,\\
\hphantom{$^\ddag$}~Fontanka 27, 191023 St.~Petersburg, Russia}
\EmailD{\href{mailto:derkach@pdmi.ras.ru}{derkach@pdmi.ras.ru}}

\Address{$^\S$~Laboratory of Theoretical Physics, JINR, Dubna, Moscow region, 141980, Russia}
\EmailD{\href{mailto:spiridon@theor.jinr.ru}{spiridon@theor.jinr.ru}}

\ArticleDates{Received November 17, 2015, in f\/inal form March 04, 2016; Published online March 11, 2016}

\Abstract{We start from known solutions of the Yang--Baxter equation with a
spectral parameter def\/ined on the tensor product of two inf\/inite-dimensional
principal series representations of the group $\mathrm{SL}(2,\mathbb{C})$ or
Faddeev's modular double. Then we describe its restriction to an irreducible
f\/inite-dimensional representation in one or both spaces.
In this way we obtain very simple explicit formulas embracing
rational and trigonometric f\/inite-dimensional solutions of the Yang--Baxter
equation. Finally, we construct these f\/inite-dimensional solutions
by means of the fusion procedure and f\/ind a nice agreement between two approaches.}


\Classification{81R50; 82B23; 33D05}

{\small \tableofcontents}

\section{Introduction}

The Yang--Baxter equation (YBE)
\begin{gather*}
\mathbb{R}_{12}(u-v) \mathbb{R}_{13}(u) \mathbb{R}_{23}(v) =
\mathbb{R}_{23}(v) \mathbb{R}_{13}(u) \mathbb{R}_{12}(u-v)
\end{gather*}
is a major tool in building the quantum integrable
systems \cite{Baxter,Jimbo,KRS81,KS81,FTT83}. It has found numerous
applications in mathematical physics and purely mathematical questions.
At the dawn of quantum inverse scattering method the
f\/inite-dimensional solutions of the YBE (when the operators~$\mathbb{R}_{ij}(u)$
are given by ordinary matrices with numerical entries depending on the spectral
parameter~$u$) attracted much attention in view of their relevance for physical
spin systems on lattices admitting a successful treatment of their
thermodynamical behavior~\cite{Baxter,Jimbo}.

Solutions of the YBE for inf\/inite-dimensional representations revealed
their importance in the integrability phenomena
emerging in quantum f\/ield theories.
An integrable spin chain with underlying $\mathrm{SL}(2,\mathbb{C})$
symmetry group and its noncompact representations naturally arises in the
high-energy behavior of quantum chromodynamics. Corresponding model was
discovered in~\cite{Lipatov:1993qn} together with an additional integral
of motion. Later, in~\cite{Lipatov:1993yb} and~\cite{Faddeev:1994zg} it
was identif\/ied with the noncompact XXX spin chain which revealed its
complete integrability (for further investigations of this model,
see \cite{DeVega:2001pu, DKM01}).

There are three increasing levels of complexity of f\/inite-dimensional
solutions of YBE described by matrices with the coef\/f\/icients
expressed in terms of the rational, trigonometric, and elliptic functions.
In the inf\/inite-dimensional setting the latter hierarchy is replaced by
solutions of YBE def\/ined as integral operators with the integrands
described by plain hypergeometric,
$q$-hypergeometric and elliptic hypergeometric functions \cite{spi:umn2}.

The notion of the modular double was introduced by Faddeev in~\cite{F99} and
noncompact representations of this algebra arise naturally in the Liouville model
studies~\cite{FKV,PT}.
The quantum dilogarithm function \cite{F95} plays an important role in
the description of these representations as well as in the Faddeev--Volkov
model~\cite{Bazhanov:2007mh, VF} and its generalization found in~\cite{spi:conm}.
The elliptic modular double extending Faddeev's double was introduced in~\cite{AA2008}.

The general solution of YBE at the elliptic level with the rank~1 symmetry algebra
was found in~\cite{DS1}.
It is based on the properties of an integral operator with an elliptic
hypergeometric kernel, the key identity for which (given by the Bailey
lemma, see, e.g., \cite{spi:umn2}) coincides with the star-triangle relation.
In~\cite{DS1,DS2} a particular f\/inite-dimensional invariant space for the
representations of the elliptic modular double has been described.

The general ${\rm R}$-operator is interesting on its own.
In the case of group $\mathrm{SL}(2,\mathbb{C})$ and the Faddeev and elliptic modular doubles
it is represented by an explicit integral operator acting on the tensor product of two functional
spaces~\cite{CD14,DKM01,DM09,DS1}.
It can be thought of as a universal object since it is expected that in some
sense it conceals {\em all} solutions of YBE, particularly, the f\/inite-dimensional
solutions. In this paper we show explicitly that, indeed, the latter solutions can
be derived as reductions of the inf\/inite-dimensional ${\rm R}$-operators
in three particular cases: the $\mathrm{SL}(2,\mathbb{C})$ group ${\rm R}$-operator~\cite{DM09},
its real form analogue associated with the ${\mathfrak{sl}}_2$-algebra and the ${\rm R}$-operator
for the Faddeev modular double, which was considered f\/irst in~\cite{BT06}
as a formal function with an operator argument.

Reductions to f\/inite-dimensional invariant subspaces
constitute a nontrivial problem.
Indeed, general inf\/inite-dimensional ${\rm R}$-matrices are given by
integral operators, but their reduction to a~f\/inite-dimensional invariant subspace in one of the tensor product spaces
should be a matrix with the entries described by dif\/ferential
or f\/inite-dif\/ference operators.

Our key results are given by the
remarkably compact formulas for reduced ${\rm R}$-operators~\eqref{redsl2}, \eqref{redsl2alg},
and~\eqref{redmod}. The former and the latter cases are determined by a pair of
integer parameters. In the $\mathrm{SL}(2,\mathbb{C})$-case~\eqref{redsl2} two integers emerge from the discretization of two spin
variables, $s$~and~$\bar s$. In the modular double case~\eqref{redmod}
the situation is qualitatively dif\/ferent, two integers emerge from the
intrinsically two-dimensional nature of
the discrete lattice for {\em one} spin variable. In the context of
univariate spectral problems such a quantization leads to the two-index orthogonality
relations which were found for the f\/irst time in the theory of elliptic
hypergeometric functions, see~\cite{spi:umn2} and references therein.
In our problem, two integers appearing in the reduction of
${\rm R}$-matrix associated with the modular double are descendants
from the similar integers existing at the elliptic level~\cite{DS1,DS2}.

It is well known that quantum integrable systems are related to $6j$-symbols
of dif\/ferent algebras. In the context of $2d$ conformal f\/ield theory
these symbols are associated with the fusion matrices and, in this setting,
the f\/inite-dimensional $6j$-symbols of the modular double with
$q$ a root of unity have been constructed in~\cite{2dcft}. Their continuous
spin generalizations have been built in~\cite{PT}. The most general discrete
$q$-$6j$-symbols of such type (with the doubling of indices) are composed out
of the product of two particular terminating~${}_{10}\varphi_9$ basic
hypergeometric series related by a modular transformation~\cite{spi:umn2}.
Their noncompact analogues associated with
the lattice model of~\cite{spi:conm} and generalizing $6j$-symbols of \cite{PT} are
easily derived as a limiting case of the elliptic
analogue of the Euler--Gauss hypergeometric function \cite{spi:umn2}.
A similar set of questions was discussed recently for the quantum algebra
 $U_q({\mathfrak{osp}}(1|2))$~\cite{PSS}.

A conventional method of constructing higher spin ${\rm L}$-operators
or the higher spin ${\rm R}$-matrices which are f\/inite-dimensional
in both spaces is the fusion procedure~\cite{KRS81,KS81}.
It is based on the fact that arbitrary
f\/inite-dimensional representation of a rank~1 algebra is contained in the decomposition
of a tensor power of the fundamental representation.
Similarly, by means of the fusion procedure one constructs higher quantized spin
solutions of YBE out of the fundamental one. In particular, a higher spin
${\rm R}$-operator, which is f\/inite-dimensional in one of the spaces,
is given by a symmetrized tensor product of several Lax operators,
and higher spin ordinary matrix solutions of YBE
are given by symmetrized tensor products of several fundamental ${\rm R}$-matrices.

There is another method of building such ${\rm R}$-operators based on the observation that
for special values of the spins (representation parameters) the principal series
representation becomes reducible and a f\/inite-dimensional irreducible representation
decouples. The general ${\rm R}$-operator does not map out of this invariant
f\/inite-dimensional subspace, so it can be restricted to this subspace and get a reduced form.
In this approach the intertwining operators of equivalent representations
of the symmetry algebras play a crucial role. They explicitly indicate
specif\/ic values of the spin when such a decoupling takes place.

In this work we elaborate both methods for the $\mathrm{SL}(2,\mathbb{C})$ group and
the modular double (the corresponding intertwining operators were constructed
in~\cite{Gelfand} and~\cite{PT}). We show explicitly that both methods yield identical
formulae embracing required f\/inite-dimensional (in one or both spaces) solutions of YBE.
Additionally, we consider a f\/inite-dimensional reduction of the ${\rm R}$-operators for
a tensor product of two Verma modules.
This is the f\/irst paper in the series dedicated to f\/inite-dimensional reductions
of known integral ${\rm R}$-operators. In the next work of this series~\cite{CDS}
such a problem was solved for the elliptic modular double.
In~\cite{CD15} new compact factorization formulae were derived for f\/inite-dimensional
${\rm R}$-matrices in several cases (for dif\/ferent forms of
factorizations, see~\cite{KhT}
and references therein). Reduction of the integral ${\rm R}$-operator for the generalized
Faddeev--Volkov model of \cite{spi:conm} is considered in~\cite{CS}.

The paper consists of two parts.
In the f\/irst part we consider $\mathrm{SL}(2,\mathbb{C})$-invariant solutions of YBE.
We begin in Section~\ref{SL2rep} with a concise review of the inf\/inite-dimensional
principal series representation of the $\mathrm{SL}(2,\mathbb{C})$ group.
In Section~\ref{RSL2} we indicate the relevant Lax operators
and the general ${\rm R}$-operator emphasizing the role of the star-triangle relation.
In Section~\ref{RedSL2} we reduce the general $\mathrm{SL}(2,\mathbb{C})$-symmetric
${\rm R}$-operator to a f\/inite-dimensional representation in one of the spaces.
In Section~\ref{sl2R} we derive an analogous reduction for the general ${\mathfrak{sl}}_2$-algebra
${\rm R}$-operator to the space of polynomials or the Verma module.

\looseness=-1
Then we proceed to the fusion. In Section~\ref{FusSL2} we formulate the fusion for
the ${\mathfrak{sl}}_2$ algebra case in a rather nonstandard fashion. We construct projectors
to the highest spin representation by means of some auxiliary spinor variables that
results in the Jordan--Schwinger realization of the ``fused'' representation.
We describe also how the fusion procedure reproduces the L-operator as well.
After that, in Section~\ref{fusSL2} we get back to the $\mathrm{SL}(2,\mathbb{C})$ group
and carry out the fusion in this case.

In the second part of the paper we consider similar questions for the modular double.
There the presentation closely follows the rational case
in order to emphasize the striking similarity between these two cases.
In Sections~\ref{qqsl2} and~\ref{3.2} we outline general structure of the
modular double and present the general ${\rm R}$-operator for it.
The corresponding reduced ${\rm R}$-matrix, which is f\/inite-dimensional in one of the
quantum spaces (or both), is derived in Section~\ref{RedMod}. Finally,
in Sections~\ref{3.4} and~\ref{fusMD} we derive f\/inite-dimensional ${\rm R}$-matrices
in the $q$-deformed cases using the fusion procedure.

\section[$\mathrm{SL}(2,\mathbb{C})$ group]{$\boldsymbol{\mathrm{SL}(2,\mathbb{C})}$ group}

\subsection{Representations of the group and the intertwining operator} \label{SL2rep}

We start with a short review of some basic well-known facts about representations
of the group $\mathrm{SL}(2,\mathbb{C})$. They are formulated in a form that
will be natural for dealing with ${\rm R}$-operators.
We outline how f\/inite-dimensional representations
decouple from inf\/inite-dimensional ones emphasizing the role of the intertwining operator.

The method of induced representations is a robust tool that
enables one to construct a number of interesting representations of a
group (see for example~\cite{Gelfand-Naimark}).
Consider representations of the group $\mathrm{SL}(2,\mathbb{C})$ realized
on the space of single-valued functions $\Phi(z,\bar{z})$ on the complex plane.
The principal series representation~\cite{Gelfand}
is parametrized by a pair of generic complex numbers~$(s,\bar{s})$
subject to the constraint~$2 (s-\bar{s}) \in \mathbb{Z}$.
We refer to them as {\it spins} in what follows.
In order to avoid misunderstanding we emphasize that $s$ and $\bar{s}$ are not
complex conjugates in general. So, this representation
$\mathrm{T}^{(s,\bar{s})}$ is given explicitly as~\cite{Gelfand}
\begin{gather}\label{tsl2}
\big[ \mathrm{T}^{(s,\bar{s})}(g) \Phi \big](z,\bar{z}) = \left(d -bz\right)^{2s}
\left(\bar{d} -\bar{b}\bar{z}\right)^{2\bar{s}}
 \Phi\left(\frac{-c+a z}{d-b z} , \frac{-\bar{c}+\bar{a}
\bar{z}}{\bar{d}-\bar{b}\bar{z}}\right),\\
 g = \left(\begin{matrix}
 a & b \\
 c & d
\end{matrix}\right) \in \mathrm{SL}(2,\mathbb{C}).\nonumber
\end{gather}
Representations of the group $\mathrm{SL}(2,\mathbb{C})$ yield representations of
the Lie algebra ${\mathfrak{sl}}(2,\mathbb{C})$ in a~standard way.
Assuming that $g$ lies in a vicinity of the identity $g = 1 + \epsilon\cdot \mathcal{E}_{ik}$,
where $\mathcal{E}_{ik}$ are traceless $2 \times 2$ matrices:
$
(\mathcal{E}_{ik})_{jl} =
\delta_{i j} \delta_{kl}-\frac{1}{2} \delta_{ik} \delta_{jl},
$
one extracts generators
$\mathrm{E}_{ik}$ and $\bar{\mathrm{E}}_{ik}$ of the Lie algebra,
\begin{gather*}
\mathrm{T}^{(s,\bar{s})}(1+\epsilon\cdot \mathcal{E}_{ik}) \Phi(z,\bar{z})=
\Phi(z,\bar{z})+\left(\epsilon\cdot
\mathrm{E}_{ik}+\bar\epsilon\cdot\bar{\mathrm{E}}_{ik}\right)
\Phi(z,\bar{z})+O\big(\epsilon^2\big).
\end{gather*}
The generators $\mathrm{E}_{ik}$, $\bar{\mathrm{E}}_{ik}$ are the
f\/irst-order dif\/ferential operators.
We arrange them in $2\times 2$ matrices $\mathrm{E}^{(s)}$
and $\bar{\mathrm{E}}^{(\bar{s})}$, which will be useful for the following considerations,
\begin{gather}\label{Egl2}
\mathrm{E}^{(s)} = \left(
\begin{matrix}
 \mathrm{E}_{11} & \mathrm{E}_{21} \\
 \mathrm{E}_{12} & \mathrm{E}_{22}
\end{matrix}%
\right) = \left(%
\begin{matrix}
 z\partial-s & -\partial \\
 z^2\partial -2s z & -z\partial+s
\end{matrix}%
\right) = \begin{pmatrix}
1 & 0 \\ z & 1
\end{pmatrix}
\begin{pmatrix}
-s-1 & -\partial \\ 0 & s
\end{pmatrix}
\begin{pmatrix}
1 & 0 \\ -z & 1
\end{pmatrix}.
\end{gather}
The substitution $z \to \bar{z}$, $\partial \to \bar{\partial}$ and $s \to \bar{s}$ in
this formula results in
the matrix $\mathrm{\bar{E}}^{(\bar{s})}$ for the generators~$\mathrm{\bar{E}}_{ik}$.

There exists an integral operator $\mathrm{W}$ which intertwines a pair of
principal series
representations $\mathrm{T}^{(s,\bar{s})}$ and $\mathrm{T}^{(-1-s,-1-\bar{s})}$
for generic complex~$s$ and~$\bar{s}$,
\begin{gather}
\label{splet}
\mathrm{W}(s,\bar{s}) \mathrm{T}^{(s,\bar{s})}(g) =
\mathrm{T}^{(-1-s,-1-\bar{s})}(g) \mathrm{W}(s,\bar{s}).
\end{gather}
We will refer to this pair as the {\it equivalent} representations.
The described intertwining relation can be equally reformulated as a set of intertwining
relations for the Lie algebra generators
\begin{gather}
\label{spletE}
\mathrm{W}(s,\bar{s}) \mathrm{E}^{(s)} =
\mathrm{E}^{(-1-s)} \mathrm{W}(s,\bar{s}), \qquad
\mathrm{W}(s,\bar{s}) \mathrm{\bar{E}}^{(\bar{s})} =
\mathrm{\bar{E}}^{(-1-\bar{s})} \mathrm{W}(s,\bar{s}).
\end{gather}
The operator $\mathrm{W}$ is def\/ined up to an overall normalization and
has the following explicit form~\cite{Gelfand}
\begin{gather} \label{dint}
\left[ \mathrm{W}(s,\bar{s}) \Phi \right](z,\bar{z}) = \text{const}
\int_{\mathbb C} d^2 x \frac{\Phi(x,\bar{x})}{(z-x)^{2s+2}(\bar{z}-\bar{x})^{2\bar{s}+2}}.
\end{gather}
Obviously this integral operator is well-def\/ined for generic values of~$s$ and $\bar{s}$
and the problems emerge for the discrete set of points $2s = n$, $2\bar{s} = \bar{n}$
with $n, \bar{n} \in \mathbb{Z}_{\geq 0}$. These special values of the spins
correspond to f\/inite-dimensional representations which we are aiming at.
That is why we would like to have a meaningful intertwining operator for
this discrete set.
In order to obtain it we note that
the expression~\eqref{dint}, considered as an analytical function of~$s$,~$\bar{s}$,
has simple poles exactly on this discrete set of (half)-integer points.
Consequently, we need to choose properly the normalization constant in~\eqref{dint}
to suppress the poles at $2s = n$, $2\bar{s} = \bar{n}$.
Further, pursuing this strategy we f\/ind the normalization constant as an
appropriate combination of the Euler gamma functions such that
the intertwining operator~\eqref{dint} becomes well-def\/ined in the case of
f\/inite-dimensional representations as well.
In order to implement the outlined program
we resort to the text-book formula for the following
complex Fourier transformation~\cite{Gelfand}
\begin{gather}\label{A}
A(\alpha,\bar\alpha) \int_{\mathbb{C}} d^2 z \frac{\mathrm{e}^{ipz+i\bar{p}\bar{z}}}
{z^{1+\alpha}\bar{z}^{1+\bar\alpha}} =
p^{\alpha}\bar{p}^{\bar\alpha} , \qquad
A(\alpha,\bar\alpha) =
\frac{i^{-|\alpha-\bar\alpha|}}{\pi}
 \frac{\Gamma\big(\frac{\alpha+\bar\alpha+|\alpha-\bar\alpha|+2}{2}\big)}
{\Gamma\big(\frac{-\alpha-\bar\alpha+|\alpha-\bar\alpha|}{2}\big)},
\end{gather}
where $\Gamma(x)$ is the Euler gamma function. One can substitute here $z=x+ iy$,
$\bar{z}=x-iy$ and pass to the integrations over~$x,y\in\mathbb{R}$.
We replace $p$ and $\bar p$ by the dif\/ferential operators, $p \to i\partial_x$ and
$\bar{p} \to i\partial_{\bar{x}}$, use the shift operator
$\mathrm{e}^{a\partial_x}f(x)=f(x+a)$, and come to the def\/inition
\begin{align}\label{kern}
\left(i\partial_z\right)^{\alpha}
\left(i\partial_{\bar{z}}\right)^{\bar\alpha}\Phi(z,\bar z) :=&
A(\alpha,\bar\alpha) \int_{\mathbb C} d^2 x
(z-x)^{-1-\alpha}(\bar{z}-\bar{x})^{-1-\bar\alpha}
\Phi(x,\bar x).
\end{align}
In order to avoid cumbersome expressions we prefer to recast this
formula to a concise form
\begin{align}\label{kern1}
\left[i\partial_z\right]^{\alpha}\Phi(z,\bar z) =&
A(\alpha) \int_{\mathbb C} d^2 x
\left[z-x\right]^{-1-\alpha}
\Phi(x,\bar x).
\end{align}
Here and in the following we prof\/it from the shorthand notation
\begin{align} \label{not1}
[z]^{\alpha} = z^{\alpha} \bar{z}^{\bar{\alpha}} , \qquad \alpha - \bar{\alpha} \in \mathbb{Z} ,
\end{align}
which unif\/ies the holomorphic and antiholomorphic sectors. Let us remind once more that~$\alpha$ and~$\bar{\alpha}$ are not assumed to be complex conjugates.
The constraint on the exponents~$\alpha$,~$\bar{\alpha}$ in~\eqref{not1} ensures that
the function $[z]^{\alpha}$ is single-valued, whereas for generic values of~$\alpha$ the holomorphic and anti-holomorphic
factors of~$[z]^{\alpha}$ taken separately have branch cuts.
Bearing in mind that the holomorphic sector is always accompanied by the antiholomorphic one
we omit the $\bar\alpha$-dependence in the $A$-factor:
$A(\alpha,\bar\alpha)\to A(\alpha)$.

Thus, if the normalization in \eqref{dint} is chosen properly, the intertwining
operator can be represented in two equivalent forms, either as a formal complex power of
the dif\/ferentiation operator
$\mathrm{W}(s,\bar{s}) = \left[i\partial_z\right]^{2s+1}$
or as a well def\/ined integral operator
\begin{gather} \label{W}
\left[ \mathrm{W}(s,\bar{s}) \Phi \right](z,\bar{z}) = \frac{(-1)^{|s-\bar{s}|}}{\pi}
 \frac{\Gamma\left(s+\bar{s}+|s-\bar{s}|+2\right)}
{\Gamma\left(-s-\bar{s}+|s-\bar{s}|-1\right)}
\int_{\mathbb C} \mathrm{d}^2 x [z-x]^{-2s-2} \Phi(x,\bar{x}).
\end{gather}
At special points $2s = n$, $2\bar{s} = \bar{n}$, $n, \bar{n} \in \mathbb{Z}_{\geq 0}$,
the integral operator turns to the dif\/ferential ope\-rator
of a f\/inite order~$\left(i\partial_z\right)^{n+1}
\left(i\partial_{\bar{z}}\right)^{\bar{n}+1}$. Let us note that for generic $s$ the holomor\-phic~$\partial_z^{2s+1}$ and
anti-holomor\-phic~$\partial_{\bar{z}}^{2\bar{s}+1}$ parts (see~\eqref{not1}) of the intertwiner $\left[i\partial_z\right]^{2s+1}$
taken separately are ill-def\/ined (working with the contour integrals with
the kernel~$(z-x)^\alpha$
one cannot f\/ind a~translationally invariant measure).
However, being taken together, they form a well-def\/ined integral operator.

Formula~(\ref{tsl2}) implies that for
special values of spins $2s = n$, $2\bar{s} = \bar{n}$ discussed above an \linebreak
$(n+1)(\bar{n}+1)$-dimensional
representation decouples from the general inf\/inite-dimensional ca\-se~\cite{Gelfand}.
Indeed, the space of polynomials
spanned by $(n+1) (\bar{n}+1)$ basis vectors
$z^{k} \bar{z}^{\bar{k}}$, where $k = 0, 1 , \ldots , n$ and $\bar{k} = 0, 1 ,
\ldots , \bar{n}$, is invariant with respect
to the action of the opera\-tors~$\mathrm{T}^{(s,\bar{s})}(g)$.
Instead of working with the separate basis vectors we prefer to deal with
a single generating function which contains all of them.
The generating function for basis vectors of this f\/inite-dimensional
representation can be chosen in the following form
\begin{gather} \label{genfunsl2}
 [z-x]^{n} = (z-x)^{n} (\bar{z} - \bar{x})^{\bar{n}},
\end{gather}
where $x$, $\bar{x}$ are some auxiliary parameters.
Indeed, expanding \eqref{genfunsl2} with respect to $x$ and $\bar{x}$
we recover all $(n+1)(\bar{n}+1)$ vectors $z^{k} \bar{z}^{\bar{k}}$,
where $k = 0, 1 , \ldots , n$ and $\bar{k} = 0, 1 , \ldots , \bar{n}$.

The decoupling of a f\/inite-dimensional representation and
the explicit expression for the ge\-ne\-rating function~\eqref{genfunsl2}
allow us to give a very natural interpretation to the situation
from the point of view of the intertwining operator.
Indeed, an immediate consequence of the def\/inition~(\ref{splet}) is that
the null-space of $\mathrm{W}(s,\bar{s})$~-- the space annihilated by the operator~--
is invariant under the action of the operators
$\mathrm{T}^{(s,\bar{s})}(g)$. Therefore, if the intertwining operator has a nontrivial null-space then
a sub-representation decouples and the corresponding invariant subspace appears.
In the case at hand, when $2s =n$ and $2\bar{s} = \bar{n}$,
the intertwining operator turns into the dif\/ferential
operator $\partial^{n+1} \bar{\partial}^{\bar{n}+1}$.

Of course this operator annihilates all
$(n+1)(\bar{n}+1)$ basis vectors $z^{k} \bar{z}^{\bar{k}}$,
where $k = 0, 1 , \ldots , n$ and $\bar{k} = 0, 1 , \ldots , \bar{n}$,
but the whole null-space of this operator is too big (it includes all harmonic functions)
and we need some additional characterization for the considered
f\/inite-dimensional subspace. Relation~(\ref{splet}) shows that
the image of the intertwining operator $\mathrm{W}(-1-s,-1-\bar{s})$ is
also invariant under the action of the operators
$\mathrm{T}^{(s,\bar{s})}(g)$.
Moreover, formula~(\ref{W}) in the considered situation
\begin{gather}
\left[ \mathrm{W}(-1-s,-1-\bar{s})
\Phi \right](z,\bar{z}) \nonumber\\
\qquad{} = \frac{(-1)^{|s-\bar{s}|}}{\pi}
 \frac{\Gamma\left(-s-\bar{s}+|s-\bar{s}|\right)}
{\Gamma\left(s+\bar{s}+|s-\bar{s}|+1\right)}
\int_{\mathbb C} d^2 x (z-x)^{2s}(\bar{z}-\bar{x})^{2\bar{s}}
\Phi(x,\bar{x}), \label{dint1}
\end{gather}
clearly shows that for special values of the spins $2s = n$ and
$2\bar{s} = \bar{n}$ discussed above the integral in the right-hand side is
equal to a polynomial with
respect to~$z$ and $\bar z$, and the image of the operator
$\mathrm{W}(-1-s,-1-\bar{s})$ (after dropping the numerical factor $\Gamma\left(-s-\bar{s}+|s-\bar{s}|\right)$
which diverges at these points) is exactly the needed f\/inite-dimensional subspace.
After all we obtain a characterization
of our f\/inite-dimensional subspace: it is the intersection of the
null-space of the intertwining operator $\mathrm{W}(s,\bar{s})$ and of the image of
the operator $\mathrm{W}(-1-s,-1-\bar{s})$ both being properly normalized for
special values of the spins $2s = n$ and $2\bar{s} = \bar{n}$.

The intertwining operator annihilates the generating function of the
f\/inite-dimensional representation \eqref{genfunsl2}, which can be seen solely from its
basic properties. The following calculation suggests this generating function
itself. The formal dif\/ferential operator form of the intertwining operators
\begin{gather*}
\mathrm{W}(s) = \left[i\partial_z\right]^{2s+1}, \qquad
\mathrm{W}(-1-s) = \left[i\partial_z\right]^{-1-2s}
\end{gather*}
formally indicates that $\mathrm{W}(-1-s)$ and $\mathrm{W}(s)$ are inverses to each other,
\begin{gather}
\mathrm{W}(s) \mathrm{W}(-1-s) = {\hbox{{1}\kern-.25em\hbox{l}}}.
\label{inv}\end{gather}
However, this inversion relation is broken for special values of the spins.
Let us rewrite the identity~\eqref{inv}
taking into account the explicit expression for kernels of the integral operators
$\mathrm{W}(-1-s)$~\eqref{dint1} and ${\hbox{{1}\kern-.25em\hbox{l}}}$, which is given by the Dirac delta-function.
In this way we f\/ind the relation
\begin{gather*}
\left[i\partial_z\right]^{2s+1} [z-x]^{2s} = (-1)^{-|s-\bar{s}|} \pi
 \frac{\Gamma\left(s+\bar{s}+|s-\bar{s}|+1\right)}
{\Gamma\left(-s-\bar{s}+|s-\bar{s}|\right)} \delta^2(z-x).
\end{gather*}
At special points $2s=n$, $2\bar{s}=\bar{n}$ the gamma-function $\Gamma\left(-s-\bar{s}+|s-\bar{s}|\right)$
has poles, and therefore
the right-hand side of the latter formula vanishes. So, one obtains
\begin{gather} \label{Wgenfunsl2}
\left[i\partial_z\right]^{n+1} [z-x]^{n} = 0, \qquad n=0,1,2,\ldots,
\end{gather}
i.e., the generating function of the f\/inite-dimensional
representation coincides with the kernel of the
intertwining operator $\mathrm{W}(-1-n/2)$ after a proper normalization.

Our calculation may seem superf\/luous since the relation \eqref{Wgenfunsl2} is evident per se.
However, we presented it here because all its basic steps remain valid after the
trigonometric (see Section~\ref{qqsl2}) and elliptic deformations (see~\cite{DS1,DS2})
of the symmetry algebra.
The deformations complicate signif\/icantly the intertwining operator and the generating
function of f\/inite-dimensional representations such that the deformed analogues of
\eqref{Wgenfunsl2} are far from being obvious and in the elliptic case they are
much more involved~\cite{CDS,DS2}.

\subsection[The general $\mathrm{SL}(2,\mathbb{C})$-invariant ${\rm R}$-operator]{The general $\boldsymbol{\mathrm{SL}(2,\mathbb{C})}$-invariant $\boldsymbol{{\rm R}}$-operator}
\label{RSL2}

Emergence of the periodic integrable spin chain with $\mathrm{SL}(2,\mathbb{C})$
symmetry in the high energy asymptotics of quantum chromodynamics
was discovered in \cite{Faddeev:1994zg, Lipatov:1993qn,Lipatov:1993yb}.
The detailed consi\-de\-ration of the corresponding formalism was performed in \cite{DeVega:2001pu, DKM01}.
In these papers the quantum-mechanical model of interest has been solved,
i.e., the relevant Baxter Q-operator has been constructed
and the separation of variables has been implemented.
The general ${\rm R}$-operator for the $\mathrm{SL}(2,\mathbb{C})$ group
has been extensively studied in the f\/irst part of \cite{DM09} as a
simplest non\-trivial example of the general $\mathrm{SL}(N,\mathbb{C})$-construction.
Here we brief\/ly outline main steps in the construction of this ${\rm R}$-operator
before proceeding to its f\/inite-dimensional reductions.

Firstly we tailor a pair of ${\rm L}$-operators out of the Lie algebra
generators $\mathrm{E}^{(s)}$, $\mathrm{\bar{E}}^{(\bar{s})}$ \eqref{Egl2}
and the spectral parameters $u$ and $\bar{u}$ which
are assumed to be restricted similar to the representation parameters,
$u-\bar{u} \in \mathbb{Z}$ \cite{DKM01,DM09},
\begin{gather} \label{Lsl2}
\mathrm{L}(u_1,u_2) = u\cdot{\hbox{{1}\kern-.25em\hbox{l}}} + \mathrm{E}^{(s)} =
\begin{pmatrix}
1 & 0 \\ z & 1
\end{pmatrix}
\begin{pmatrix}
u_1 & -\partial \\ 0 & u_2
\end{pmatrix}
\begin{pmatrix}
1 & 0 \\ -z & 1
\end{pmatrix},
\\
 \label{barLsl2}
\bar{\mathrm{L}}(\bar{u}_1,\bar{u}_2) =
\bar{u}\cdot{\hbox{{1}\kern-.25em\hbox{l}}} + \mathrm{\bar{E}}^{(\bar{s})} =
\begin{pmatrix}
1 & 0 \\ \bar{z} & 1
\end{pmatrix}
\begin{pmatrix}
\bar{u}_1 & -\bar{\partial} \\ 0 & \bar{u}_2
\end{pmatrix}
\begin{pmatrix}
1 & 0 \\ -\bar{z} & 1
\end{pmatrix} .
\end{gather}
Here we use the convenient shorthand notation
\begin{gather} \label{u}
u_1 = u - s - 1 ,\qquad u_2 = u + s ,\qquad \bar{u}_1 = \bar{u} - \bar{s} - 1 ,\qquad \bar{u}_2 = \bar{u} + \bar{s}.
\end{gather}
Each of the ${\rm L}$-operators \eqref{Lsl2}, \eqref{barLsl2}
respects the $\mathrm{RLL}$-relation with Yang's $4 \times 4$ ${\rm R}$-matrix,
\begin{gather}
 \mathrm{R}_{ab,ef}(u-v) \mathrm{L}_{ec}(u)
\mathrm{L}_{fd}(v) = \mathrm{L}_{bf}(v)
\mathrm{L}_{ae}(u) \mathrm{R}_{ef,cd}(u-v), \label{RLLYang} \\ \notag
 \mathrm{R}_{ab,ef}(\bar{u}-\bar{v}) \bar{\mathrm{L}}_{ec}(\bar{u})
\bar{\mathrm{L}}_{fd}(\bar{v}) = \bar{\mathrm{L}}_{bf}(\bar{v})
\bar{\mathrm{L}}_{ae}(\bar{u}) \mathrm{R}_{ef,cd}(\bar{u}-\bar{v}),
\end{gather}
where $a,b, \ldots = 1,2$, and the summation over repeated indices is
assumed, $\mathrm{R}_{ab,cd}(u) = u
\cdot\delta_{ac} \delta_{bd}+\delta_{ad} \delta_{bc} $ (cf.~\eqref{Yang}).
The described relations supplemented by the commutativity condition
$[ \mathrm{L}(u) , \bar{\mathrm{L}}(\bar{v}) ]=0$ are equivalent to the set of
commutation relations for the Lie algebra generators of~$\mathrm{SL}(2,\mathbb{C})$.

The ${\rm L}$-operators \eqref{Lsl2}, \eqref{barLsl2} respect simultaneously
another $\mathrm{RLL}$-relation with some {\it general} ${\rm R}$-operator
which intertwines the co-product of ${\rm L}$-operators in the pair of quantum spaces
\begin{gather} \label{RLLsl2}
\mathrm{R}_{12}(u-v,\bar{u}-\bar{v}) \mathrm{L}_1(u_1,u_2)
 \mathrm{L}_2(v_1,v_2)
= \mathrm{L}_1(v_1,v_2) \mathrm{L}_2(u_1,u_2) \mathrm{R}_{12}(u-v,\bar{u}-\bar{v}) ,
\\
\mathrm{R}_{12}(u-v,\bar{u}-\bar{v}) \bar{\mathrm{L}}_1(\bar{u}_1,\bar{u}_2)
 \bar{\mathrm{L}}_2(\bar{v}_1,\bar{v}_2)
= \bar{\mathrm{L}}_1(\bar{v}_1,\bar{v}_2)
 \bar{\mathrm{L}}_2(\bar{u}_1,\bar{u}_2)
 \mathrm{R}_{12}(u-v,\bar{u}-\bar{v}) , \label{RLLsl2'}
\end{gather}
where parameters $u_1$ and $u_2$
are def\/ined in \eqref{u}, and $v_1,v_2$ are analogous linear combinations of $v$ and $\ell$,
\begin{gather*}
v_1 = v - \ell - 1 ,\qquad v_2 = v + \ell ,\qquad \bar{v}_1 = \bar{v} - \bar{\ell} - 1 ,\qquad \bar{v}_2 = \bar{v} + \bar{\ell} .
\end{gather*}
The lower indices of $\mathrm{R}_{12}$ and $\mathrm{L}_1$, $\mathrm{L}_2$
denote quantum spaces on which the operators act non-trivially.
The ${\rm L}$-operators are multiplied as conventional $2\times 2$ matrices and the ${\rm R}$-operator acts as an identity operator on the auxiliary $2$-dimensional spaces of ${\rm L}$-operators,
but it acts non-trivially on the tensor product of two inf\/inite-dimensional representations:
the f\/irst representation is specif\/ied by the spins~$s$,~$\bar{s}$ and it is realized on the
functions of variables~$z_1$,~$\bar{z}_1$, the second representation is specif\/ied
by the spins~$\ell$,~$\bar{\ell}$ and it is realized on the functions of variables~$z_2$,~$\bar{z}_2$.
In~\eqref{RLLsl2},~\eqref{RLLsl2'} we drop dependencies of the ${\rm R}$-operator on the representation parameters.
The full-f\/ledged notation would be $\mathrm{R}(u-v,\bar{u}-\bar{v}\,|\,s,\bar{s},\ell,\bar{\ell})$.

Note that the ${\rm R}$-operator serves for both ${\rm L}$-operators,
i.e., it is not just the holomorphic or anti-holomorphic object,
as opposed to the ${\rm L}$-operators~\eqref{Lsl2},~\eqref{barLsl2}.
In the following we frequently omit the dependence of the ${\rm R}$-operator (and other intertwining operators)
on the anti-holomorphic parameters denoting it~$\mathrm{R}(u)$.
The ${\rm R}$-operator is invariant with respect to the $\mathrm{SL}(2,\mathbb{C})$ group, i.e.,
it commutes with the co-product of ${\mathfrak{sl}}(2,\mathbb{C})$ generators
\begin{gather*}
\big[ \mathrm{R}_{12}(u,\bar{u}) , \mathrm{E}^{(s)}_1 + \mathrm{E}^{(\ell)}_2 \big] = 0 , \qquad
\big[ \mathrm{R}_{12}(u,\bar{u}) , \mathrm{\bar{E}}^{(\bar{s})}_1
+ \mathrm{\bar{E}}^{(\bar{\ell})}_2 \big] = 0,
\end{gather*}
which follows immediately from the $\mathrm{R}\mathrm{L}\mathrm{L}$-relations~\eqref{RLLsl2} and~\eqref{RLLsl2'}.

Apart from the $\mathrm{RLL}$-relations \eqref{RLLsl2}, \eqref{RLLsl2'} the general
${\rm R}$-operator satisf\/ies the YBE
\begin{gather} \label{YB}
\mathrm{R}_{23}(u-v,\bar{u}-\bar{v}) \mathrm{R}_{12}(u,\bar{u}) \mathrm{R}_{23}(v,\bar{v}) =
\mathrm{R}_{12}(v,\bar{v}) \mathrm{R}_{23}(u,\bar{u})
\mathrm{R}_{12}(u-v,\bar{u}-\bar{v}) ,
\end{gather}
where both sides are endomorphisms on the tensor product of three inf\/inite-dimensional
spaces realizing arbitrary principal series representations of~$\mathrm{SL}(2,\mathbb{C})$.

In \cite{DKM01,DM09} an integral operator solution of the intertwining
relations~\eqref{RLLsl2} and \eqref{RLLsl2'} was found, which solves simultaneously YBE \eqref{YB}.
The construction naturally gives to this general ${\rm R}$-operator several
factorized forms related to an integral operator realization of the generators
of symmetric group $\mathfrak{S}_4$ \cite{DM09}.
Here we do not go into details of this formalism and just indicate the
factorization which is appropriate for our current purposes.
The ${\rm R}$-operator can be represented as a product of four
{\it elementary intertwining} operators \cite{DM09}
\begin{gather}
\mathrm{R}_{12}(u-v,\bar{u}-\bar{v}) =[z_{12}]^{u_2-v_1}\left[i\partial_{2}\right]^{u_1-v_1}
\left[i\partial_{1}\right]^{u_2-v_2}[z_{12}]^{u_1-v_2} \label{R},
\end{gather}
where we assume the shorthand notation $z_{ij} = z_i - z_j$ and \eqref{not1}.
Taking into account \eqref{kern1} one can rewrite \eqref{R} explicitly as an integral operator.
The notation \eqref{not1} implies that the ${\rm R}$-operator consists of the holomorphic and anti-holomorphic parts
which, being taken separately, are ill-def\/ined for generic spectral and representation parameters.
The merge of holomorphic and antiholomorphic parts yields a well-def\/ined integral operator.

Formula \eqref{R} plays a crucial role in the subsequent discussion.
It admits deformations \cite{Derkachov:2007gr} leading to
${\rm R}$-operators for the modular double \cite{CD14}
and the elliptic modular double \cite{DS1}.

The expression \eqref{R} may seem rather unusual. In \cite{DM09} it was shown that
the holomorphic part of the ${\rm R}$-operator \eqref{R}, being restricted to the space
of polynomials, coincides with the familiar ${\rm R}$-operator constructed
in \cite{KRS81,FTT83} in the form of the beta-function depending on the ``square root''
of the Casimir operator.
However, the form \eqref{R} does not demand extra information about the structure of
tensor products and corresponding Clebsch--Gordan coef\/f\/icients.
Furthermore, we will show that the integral ${\rm R}$-operator \eqref{R}
contains f\/inite-dimensional solutions of the Yang--Baxter relation as well \eqref{YB}.

The elementary intertwining operators appearing in \eqref{R}
fulf\/ill the following operator relations
\begin{gather}\label{str}
\left[i\partial_{k}\right]^{a}
[z_{12}]^{a+b} \left[i\partial_{k}\right]^{b} = [z_{12}]^{b} \left[i\partial_{k}\right]^{a+b} [z_{12}]^{a}, \qquad k=1,2 .
\end{gather}
These formulae have a remarkable interpretation in terms of
the Coxeter relations of the symmetric group $\mathfrak{S}_4$ \cite{Derkachov:2007gr,DM09}.
Using~(\ref{str}) one can easily prove that the ${\rm R}$-operator~(\ref{R}) respects
the YBE~(\ref{YB}). The operator factors in \eqref{R} are called intertwiners
because they satisfy the equations
\begin{gather} \nonumber
\left[i\partial_{1}\right]^{u_2-u_1} \mathrm{L}_1(u_1,u_2) =
\mathrm{L}_1(u_2,u_1) \left[i\partial_{1}\right]^{u_2-u_1}, \\ \label{rel1}
\left[i\partial_{2}\right]^{v_2-v_1} \mathrm{L}_2(v_1,v_2) =
\mathrm{L}_2(v_2,v_1) \left[i\partial_{2}\right]^{v_2-v_1}, \\ \label{rel2}
[z_{12}]^{u_1-v_2} \mathrm{L}_1(u_1,u_2) \mathrm{L}_2(v_1,v_2)=
\mathrm{L}_1(v_2,u_2) \mathrm{L}_2(v_1,u_1) [z_{12}]^{u_1-v_2},
\end{gather}
and similar ones with $\mathrm{L}$ substituted by $\mathrm{\bar{L}}$.
Here the operators $\left[i\partial_{k}\right]^{a}$ and $[z_{12}]^a$
act on each matrix element of the matrices $\mathrm{L}_k$ entrywise, i.e.,
they should be considered as $2\times 2$ diagonal matrices proportional to
the unit matrix. Moreover, the latter relations f\/ix uniquely
(up to a normalization) the elementary intertwining operators.
Note that $\left[i\partial_{1}\right]^{u_2-u_1} = \left[i\partial_{1}\right]^{2s+1}$
and $\left[i\partial_{2}\right]^{v_2-v_1} = \left[i\partial_{2}\right]^{2\ell+1}$ are
the intertwining operators of the equivalent representations~(\ref{splet})
for the f\/irst and second spaces, respectively.
The equalities \eqref{rel1} are identical to the def\/ining relations~(\ref{splet}) of the intertwining operator $\mathrm{W}$.
Applying several times \eqref{rel1} and \eqref{rel2} one can easily check that
the composite ${\rm R}$-operator~(\ref{R}) obeys the $\mathrm{RLL}$-relations~(\ref{RLLsl2}) and (\ref{RLLsl2'}).

The identities~(\ref{str}) are equivalent
to the famous star-triangle relation which can be
represented in the following three equivalent forms:
\begin{enumerate}\itemsep=0pt
\item[1)] as an integral identity~\cite{DKM01, Vas04}
\begin{gather} \int_{\mathbb C} d^2
w \frac{1}{[z-w]^\alpha [w-x]^\beta [w-y]^\gamma} \nonumber\\
\qquad {}=
\frac{A(-\beta)}{A(\alpha-1)A(\gamma-1)}
\frac{1}{[z-x]^{1-\gamma} [z-y]^{1-\beta} [y-x]^{1-\alpha}}, \label{uniq}
\end{gather}
provided that the exponents respect the uniqueness conditions
\begin{gather*}
\alpha+\beta+\gamma=\bar \alpha+\bar\beta+\bar\gamma=2 ;
\end{gather*}
\item[2)] as a particular point in the image of the operator
$\left[i\partial_z\right]^{\alpha-1}$ (with the same restriction on the exponents as before)
\begin{gather}\label{uniq1}
\left[i\partial_z\right]^{\alpha-1}\left(\frac{1}{[z-x]^{\beta}[z-y]^{\gamma}}\right) =
\frac{A(-\beta)}{A(\gamma-1)}
 \frac{1}{[z-x]^{1-\gamma}[z-y]^{1-\beta}[y-x]^{1-\alpha}} ;
\end{gather}
\item[3)] or as a pseudo-dif\/ferential operators identity~\cite{Is03}
\begin{gather}
[i\partial_z]^{\alpha}\cdot
[z]^{\alpha+\beta}\cdot[i\partial_z]^{\beta} = [z]^{\beta}\cdot
[i\partial_z]^{\alpha+\beta}\cdot[z]^{\alpha}. \label{unique}
\end{gather}
\end{enumerate}

\subsection[Finite-dimensional reductions of the general ${\rm R}$-operator]{Finite-dimensional reductions of the general $\boldsymbol{{\rm R}}$-operator}
\label{RedSL2}

Now we reduce the ${\rm R}$-operator \eqref{R} to f\/inite-dimensional representations in its f\/irst space.
The principal possibility of this reduction is based on the following relation
\begin{gather}
\label{base}
\left[i\partial_{1}\right]^{u_2-u_1} \mathbb{R}_{12}(u_1,u_2\,|\,v_1,v_2) =
\mathbb{R}_{12}(u_2,u_1\,|\,v_1,v_2) \left[i\partial_{1}\right]^{u_2-u_1},
\end{gather}
where we use the $\mathbb{R}$-operator $\mathbb{R}_{12}: = \mathrm{P}_{12}\mathrm{R}_{12}$
with $\mathrm{P}_{12}$ -- a permutation operator, $\mathrm{P}_{12} \Psi(z_1,z_2) =
\Psi(z_2,z_1) \mathrm{P}_{12}$. Relation \eqref{base} can be proved using the
identity~(\ref{str}) and it shows that both, the null-space
of the intertwining operator $\left[i\partial_{1}\right]^{2s+1}$ and the image of
the intertwining opera\-tor~$\left[i\partial_{1}\right]^{-2s-1}$, are mapped onto
themselves by our ${\rm R}$-matrix~$\mathbb{R}_{12}$. Therefore, if we f\/ind invariant f\/inite-dimensional
subspaces of the latter spaces they will be invariant with respect to the action
of ${\rm R}$-operator itself.

We take the function $[z_{13}]^{2s}\Phi(z_2,\bar{z}_2)$, where $2s=u_2-u_1-1$
and $\Phi(z_2,\bar{z}_2)$ is an arbitrary function, and act upon it by the ${\rm R}$-operator.
We break down the calculation to several steps according to the factorized form \eqref{R} of the ${\rm R}$-operator.
At the end of calculation we choose $2s = n$, $2\bar{s} = \bar{n}$ with $n,\bar{n} \in \mathbb{Z}_{\geq 0}$
such that $[z_{13}]^{2s}$ turns into the generating function of the f\/inite-dimensional representation in the f\/irst space~\eqref{genfunsl2}
with an auxiliary parameter~$z_3$. However, for a while we assume the spin $s$ to be generic.

Using formula~(\ref{uniq1}) we implement the f\/irst step.

We act by
the f\/irst two factors $\left[i\partial_{1}\right]^{u_2-v_2}[z_{12}]^{u_1-v_2}$
of the ${\rm R}$-operator \eqref{R} and f\/ind
\begin{gather}
\left[i\partial_{1}\right]^{u_2-v_2}[z_{12}]^{u_1-v_2} [z_{13}]^{2s} \Phi(z_2,\bar{z}_2)\nonumber\\
\qquad{}
= \frac{A(u_1-v_2)}{A(u_1-u_2)} \cdot [z_{12}]^{u_1-u_2} [z_{13}]^{v_2-u_1-1} [z_{23}]^{u_2-v_2}\Phi(z_2,\bar{z}_2) . \label{step1}
\end{gather}
In order to apply the third factor $[i\partial_2]^{u_1-v_1}$ of the ${\rm R}$-operator \eqref{R} we resort to the relation
\begin{gather*}
[i\partial_2]^{u_1-v_1} [z_{12}]^{u_1-u_2} [z_{23}]^{u_2-v_2} \Phi(z_2,\bar{z}_2)\\
\qquad{} =
 \frac{A(u_1-v_1)}{A(u_2-u_1-1)} \cdot [i \partial_1]^{u_2-u_1-1}
[z_{12}]^{v_1-u_1-1} [z_{13}]^{u_2-v_2} \Phi(z_1,\bar{z}_1),
\end{gather*}
which follows immediately from the integral representation~\eqref{kern1}
for $\left[i\partial_z\right]^{\alpha}$.
A merit of the previous formula is that we traded the integral operator
$[i\partial_2]^{u_1-v_1}$ for $[i \partial_1]^{2s}$, which becomes just a dif\/ferential
operator for $2s = n$ and $2\bar{s} = \bar{n}$. Incorporating into the latter formula
the inert factors from \eqref{step1} and the last factor $[z_{12}]^{u_2-v_1}$
of the ${\rm R}$-operator~\eqref{R}, we f\/ind
\begin{gather}
\mathrm{R}_{12}(u_1,u_2\,|\,v_1,v_2) [z_{13}]^{u_2-u_1-1} \Phi(z_2,\bar{z}_2) =
\frac{A(u_1-v_2)}{A(u_1-u_2)}\frac{A(u_1-v_1)}{A(u_2-u_1-1)} \nonumber
\\
 \qquad {}\times [z_{12}]^{u_2-v_1} [z_{13}]^{v_2-u_1-1}
[i \partial_1]^{u_2-u_1-1} [z_{12}]^{v_1-u_1-1} [z_{13}]^{u_2-v_2} \Phi(z_1,\bar{z}_1).\label{redsl2'}
\end{gather}
In order to polish the latter formula
we denote $z_3 = x$ like in \eqref{genfunsl2} and rewrite \eqref{redsl2'} in terms of the representation parameters.
Also we prefer to replace the ${\rm R}$-operator by
$\mathbb{R}_{12} = \mathrm{P}_{12}\mathrm{R}_{12}$.

Thus the general ${\rm R}$-operator for the $\mathrm{SL}(2,\mathbb{C})$ group
acting in the tensor product of two inf\/inite-dimensional representation
spaces with spins $s$, $\bar{s}$ and $\ell$, $\bar{\ell}$ can be reduced to a
f\/inite-dimensional subspace in the f\/irst space if $2s = n$, $2\bar{s} =
\bar{n}$ ($n, \bar{n} \in \mathbb{Z}_{\geq 0}$). We have the following formula
\begin{gather}
\mathbb{R}_{12}\big(u\,|\,{\tfrac{n}{2}},
{\tfrac{\bar{n}}{2}},\ell,\bar{\ell}\big) [z_1-x]^n \Phi(z_2,\bar{z}_2) \nonumber \\
\qquad{}
= c \cdot
[z_2-x]^{-u+\frac{n}{2}+\ell} [z_{12}]^{u+\frac{n}{2}+\ell+1}
[\partial_{z_2}]^n
 [z_{12}]^{-u + \frac{n}{2} - \ell - 1}
[z_2-x]^{u + \frac{n}{2} - \ell} \Phi(z_2, \bar{z}_2),
\label{redsl2}
\end{gather}
where the normalization factor is
\begin{gather*}
c = (-1)^{n+\bar{n}} \frac{A(u - \frac{n}{2} + \ell)}{A(- u + \frac{n}{2} + \ell)}.
\end{gather*}
The latter formula gives a number of solutions of the YBE~\eqref{YB} which are
endomorphisms on the tensor product of an $(n+1)(\bar{n}+1)$-dimensional
and an inf\/inite-dimensional spaces.

We consider formula \eqref{redsl2} as one of the main results of this paper.
It gives a concise expression for known
higher spin ${\rm R}$-operators. They are ``mixed'' objects in a sense that
they are def\/ined on the tensor product of f\/inite-dimensional and inf\/inite-dimensional representations. In addition
they can be considered as generalizations of the ${\rm L}$-operators from the
fundamental to arbitrary f\/inite-dimensional representations.
Moreover, the formula \eqref{redsl2} produces all such solutions of the YBE
related to the principal series representation.
Its analogue for the modular double is derived in Section~\ref{RedMod}
and the elliptic modular double case is considered in \cite{CDS}.

In order to get accustomed to the reduction formula \eqref{redsl2} let us consider a simple example.
One can easily recover the holomorphic ${\rm L}$-operator \eqref{Lsl2}
substituting $(n,\bar{n}) = (1,0)$ in \eqref{redsl2}
and choosing the basis in the space $\mathbb{C}^2$ of the fundamental representation as
$\mathbf{e}_1 = -z_1$, $\mathbf{e}_2 = 1$.
Then
\begin{gather}
\mathbb{R}_{12}\big(u-\tfrac{1}{2}\, |\, \tfrac{1}{2},\ell\big) \mathbf{e}_1 = c \cdot \big[ \mathbf{e}_1 (z_2 \partial_2 - \ell + u)
+ \mathbf{e}_2 \big(z_2^2 \partial_2 - 2 \ell z_2\big) \big] , \\
\mathbb{R}_{12}\big(u-\tfrac{1}{2}\, |\, \tfrac{1}{2},\ell\big) \mathbf{e}_2 = c\cdot \big[ \mathbf{e}_1 (-\partial_2)
+ \mathbf{e}_2 (u + \ell - z_2 \partial_2) \big].
\end{gather}
Consequently the restriction of $\mathbb{R}_{12}(u-\tfrac{1}{2}\,|\,\tfrac{1}{2},\ell)$
to $\mathbb{C}^2$ in the f\/irst factor takes the matrix form
\begin{gather} \label{LaxNonFact}
\mathrm{L}(u) =
\begin{pmatrix}
u - \ell + z \partial & -\partial \\
z^2 \partial - 2 \ell z & u + \ell - z\partial
\end{pmatrix}
\end{gather}
and coincides with the holomorphic ${\rm L}$-operator~\eqref{Lsl2}. Analogously taking $(n,\bar{n}) = (0,1)$
we recover the anti-holomorphic $\bar{\mathrm{L}}$-operator \eqref{barLsl2}.

Besides the ${\rm L}$-operator, the formula \eqref{redsl2} reproduces all its
higher-spin generalizations.
 Simultaneously, it produces ${\rm R}$-matrices
described by plain f\/inite-dimensional matrices in both spaces.
Indeed, substituting in~\eqref{redsl2} the
generating function~\eqref{genfunsl2} of the f\/inite-dimensional $(m+1)(\bar{m}+1)$-dimensional
representation in the second space, we f\/ind a solution of the YBE~\eqref{YB} for the
spins $\frac{n}{2}$, $\frac{\bar{n}}{2}$ and $\frac{m}{2}$, $\frac{\bar{m}}{2}$
in the f\/irst and second spaces, respectively,
\begin{gather}
\mathbb{R}_{12}\big(u\,|\, \tfrac{n}{2},
\tfrac{\bar{n}}{2},
\tfrac{m}{2},\tfrac{\bar{m}}{2}\big)
 [z_1 - x]^n [z_2 - y]^m \label{redRsl2} \\
\qquad{}= c \cdot
[z_2 - x]^{-u+\frac{n}{2}+\ell} [z_{12}]^{u+\frac{n}{2}+\frac{m}{2}+1}
[\partial_{z_2}]^n
 [z_{12}]^{-u + \frac{n}{2} - \frac{m}{2} - 1} [z_2 - x]^{u + \frac{n}{2} - \frac{m}{2}} [z_2 - y]^m.
\nonumber
\end{gather}
Expanding both sides of this relation in auxiliary parameters $x$, $\bar{x}$, $y$, $\bar{y}$
one can rewrite it in a~form of a~square matrix with $(n+1)(\bar{n}+1)(m+1)(\bar{m}+1)$
rows (or columns).
The compact formula \eqref{redRsl2} produces all its entries.
In particular, taking the fundamental representation in both spaces $n = m = 1$, $\bar{n} = \bar{m} = 0$
we reproduce Yang's ${\rm R}$-matrix (cf.~\eqref{Yang}).

\subsection{Verma module reduction}
\label{sl2R}

In this section we slightly digress from the discussion of the group $\mathrm{SL}(2,\mathbb{C})$ and outline
how ${\mathfrak{sl}}_2$-symmetric f\/inite-dimensional solutions of the YBE arise from the inf\/inite-dimensional ones.
Si\-mi\-lar to the previous considerations this approach yields a concise expression for f\/inite-dimensional solutions
that may f\/ind various applications.
Since the corresponding calculations are essentially based on ideas explained
above we will limit ourselves to the statement of the results.

Although the ${\mathfrak{sl}}_2$ algebra is ``a half'' of the Lie algebra of the group $\mathrm{SL}(2,\mathbb{C})$,
it requires a special treatment. We deal with a functional representation of the
${\mathfrak{sl}}_2$-algebra in the space
of polynomials of one complex variable~$\mathbb{C}[z]$. Fixing a generic complex number $s \in \mathbb{C}$
and representing the algebra generators by the f\/irst order dif\/ferential operators given in~\eqref{Egl2}
we endow $\mathbb{C}[z]$ with a structure of the Verma module. For generic value of~$s$ the module is
an inf\/inite-dimensional space with the basis $\{1,z,z^2,\ldots\}$ and there are no invariant
subspaces, i.e., the representation is irreducible. Invariant subspaces arise for
the discrete set of
spin values $2s = n$, $n \in \mathbb{Z}_{\geq 0}$. The corresponding $(n+1)$-dimensional
representation is irreducible and it is realized on the submodule with the basis
$\{1,z,\ldots,z^n\}$.

Since the ${\mathfrak{sl}}_2$ generators are holomorphic, we have a single holomorphic ${\rm L}$-operator given in~\eqref{Lsl2}.
Now only the holomorphic spectral parameter~$u$ is present.
The general ${\rm R}$-operator $\mathrm{R}(u\,|\,s,\ell)$ is def\/ined on the tensor product of
two Verma modules with the spins~$s$ and~$\ell$.
It has to satisfy holomorphic analogues of the $\mathrm{RLL}$-relation~\eqref{RLLsl2}
and of the YBE~\eqref{YB}.

The general ${\rm R}$-operator \eqref{R} for $\mathrm{SL}(2,\mathbb{C})$ group is well def\/ined
due to its non-analyticity, in other words, due to the presence
of holomorphic and antiholomorphic parts.
We cannot get the general ${\rm R}$-operator for ${\mathfrak{sl}}_2$ (which has to be
holomorphic) by crossing out the anti-holomorphic part of \eqref{R}. Anyway, the
holomorphic $\mathrm{RLL}$-relation \eqref{RLLsl2} can be solved \cite{DM09}
in terms of a~well-def\/ined operator on $\mathbb{C}[z_1] \otimes \mathbb{C}[z_2]$ which takes
the following factorized form,
\begin{gather} \label{Rsl2R}
\mathrm{R}_{12}(u\,|\,s,\ell)=
\frac{\Gamma(z_{21}\partial_2-2s)}{\Gamma(z_{21}\partial_2 -u -s - \ell)}
\frac{\Gamma(z_{12}\partial_1 + u -s - \ell)}{\Gamma(z_{12}\partial_1-2s)},
\end{gather}
where ratios of the operator-valued gamma functions are def\/ined with
the help of the integral representation for Euler's beta-function
\begin{gather*}
\frac{\Gamma(z_{12}\partial_1 +a)}{\Gamma(z_{12}\partial_1+b)}\Phi(z_1,z_2):=
\frac{1}{\Gamma(b-a)}\int_0^1 d\alpha \alpha^{a-1}(1-\alpha)^{b-a-1}
\Phi(\alpha z_1+(1-\alpha)z_2,z_2).
\end{gather*}
This ${\rm R}$-operator satisf\/ies the holomorphic analogue of YBE \eqref{YB} as well.
As we remarked in Section~\ref{RSL2} the operator
\eqref{Rsl2R} coincides with the one found
in \cite{KRS81,FTT83} in the early days of the quantum inverse scattering method
in spite of the fact that they look completely dif\/ferent.

For $2s = n$, $n \in \mathbb{Z}_{\geq 0}$, the general ${\rm R}$-operator \eqref{Rsl2R}
can be restricted to an $(n+1)$-dimensional representation in the f\/irst space.
Taking into account permutation of the pair of tensor factors,
$\mathbb{R}_{12} = \mathrm{P}_{12}\mathrm{R}_{12} $, one can show that
the restricted ${\rm R}$-operator acquires a concise form
\begin{gather}
\mathbb{R}_{12}\big(u\,|\, \tfrac{n}{2},\ell\big) (z_1-x)^n \Phi(z_2)\nonumber\\
\qquad{} = c \cdot
(z_2-x)^{-u+\frac{n}{2}+\ell} z_{12}^{u+\frac{n}{2}+\ell+1}
\partial_{z_2}^n
 z_{12}^{-u + \frac{n}{2} - \ell - 1}
(z_2-x)^{u + \frac{n}{2} - \ell} \Phi(z_2),
\label{redsl2alg}
\end{gather}
where the normalization factor is
\begin{gather*}
c = (-1)^{n+1} \frac{\Gamma(-\ell-\frac{n}{2}-u)}{\Gamma(-\ell+\frac{n}{2}-u)}.
\end{gather*}
Formula \eqref{redsl2alg} is completely analogous to the $\mathrm{SL}(2,\mathbb{C})$ reduction
formula \eqref{redsl2}.
Expanding both sides of \eqref{redsl2alg} with respect to an auxiliary parameter
$x$ one recovers an $(n+1)\times(n+1)$-matrix whose entries are the $n$-th order dif\/ferential
operators with polynomial coef\/f\/icients in spectral parameter $u$ of degree $n$ (or lower).

In \cite{CDKK11} the Lax operator has been recovered from the general
${\rm R}$-operator by means of a quite bulky calculation.
Formula \eqref{redsl2alg} provides considerable simplif\/ication of that result
generalizing it to the higher-spin analogues of the rational Lax operator.

In order to illustrate the power of the formula \eqref{redsl2alg} we
present below the ${\rm R}$-operator for the spin $1$ representation in the f\/irst space.
In the basis $\mathbf{e}_1 = 1$, $\mathbf{e}_2 = z_1$, $\mathbf{e}_3 = z^2_1$
of the $3$-dimensional space, the $\mathbb{R}(u\,|\,1,\ell)$-operator takes the matrix
form (we change notation $z_2 \to z$)
\begin{gather*}
\begin{pmatrix}
\scriptstyle (u + \ell)(u+\ell+1) - 2 (u+\ell) z \partial + z^2 \partial^2\!\! & \scriptstyle 2 \ell (u+\ell) z - (u+3\ell-1) z^2 \partial + z^3 \partial^2
& \scriptstyle 2\ell(2\ell-1) z^2 + 2(1-2\ell)z^3\partial + z^4 \partial^2 \\
\scriptstyle 2(u+ \ell) \partial - 2 z \partial^2 & \scriptstyle (u+\ell)(u-\ell+1) + 2(2\ell-1)z\partial - 2 z^2 \partial^2\!\!
& \scriptstyle 4 \ell(u-\ell+1) z - 2(u-3\ell+2) z^2 \partial - 2 z^3 \partial^2 \\
\scriptstyle \partial^2 & \scriptstyle (u-\ell+1)\partial +z\partial^2
& \scriptstyle (u-\ell)(u-\ell+1) + 2(u - \ell +1) z \partial + z^2 \partial^2
\end{pmatrix}.
\end{gather*}
Conventional methods demand laborious calculations to reproduce
this complicated matrix. In our case the result
follows immediately from the formula \eqref{redsl2alg}.
An explicit matrix factorization formula for the operator
$\mathbb{R}_{12}(u\,|\,\frac{n}{2},\ell)$ \eqref{redsl2alg} generalizing
factorization of the ${\rm L}$-operator \eqref{Lsl2} was derived in the followup paper~\cite{CD15}.
E.g., the $\mathbb{R}(u\,|\,1,\ell)$-operator given above factorizes to a~product of f\/ive more
elementary $3 \times 3$ matrices: two lower-triangular, two diagonal
and one upper-triangular matrix.

\subsection{Fusion, symbols and the Jordan--Schwinger representation}
\label{FusSL2}
The standard procedure for constructing f\/inite-dimensional higher-spin ${\rm R}$-operators out of
the fundamental one is the {\it fusion procedure}~\cite{KRS81,KS81}. Firstly, we remind how it works
in the case of the symmetry algebra ${\mathfrak{sl}}_2$ using
a formulation convenient for us. Then in the next section we straightforwardly
extend it to the case of the $\mathrm{SL}(2,\mathbb{C})$ group and show that
the reduction formula \eqref{redsl2} is in line with the fusion construction.

For the rank one symmetry algebras underlying an integrable system
the recipe of~\cite{KRS81,KS81} looks as follows. One forms an {\it inhomogeneous} monodromy matrix
$\mathrm{T}_{i_1\ldots i_n}^{j_1 \ldots j_n}$
out of ${\rm L}$-opera\-tors~$\mathrm{L}^{j}_{i}$ multiplying them as operators in quantum space and
taking tensor products of the auxiliary space $\mathbb{C}^2$,
and then symmetrizes the monodromy matrix over the spinor indices.
The parameters of inhomogeneity have to be adjusted in a proper way.
The result $\mathrm{T}_{(i_1\ldots i_n)}^{(j_1 \ldots j_n)}$
is an ${\rm R}$-operator which has a higher-spin auxiliary space and solves the YBE.
Thus constructing higher-spin ${\rm R}$-operators one has to deal with $\operatorname{Sym} \bigl(\mathbb{C}^2\bigr)^{ \otimes n}$
which is a space of symmetric tensors with a number of spinor indices $\Psi_{(i_1 \ldots i_n)}$.
The usual matrix-like action of operators has the form
\begin{gather} \label{TPsi}
\left[\mathrm{T} \Psi\right]_{(i_1\ldots i_n)}
= \mathrm{T}_{(i_1\ldots i_n)}^{(j_1 \ldots j_n)} \Psi_{(j_1 \ldots j_n)},
\end{gather}
where the summation over repeated indices is assumed.
We prefer not to deal with a multitude of spinor indices. Instead we introduce
auxiliary spinors $\lambda=(\lambda_1,\lambda_2)$, $\mu=(\mu_1,\mu_2)$
and contract them with the tensors
\begin{gather} \label{Tlam}
\lambda_{i_1} \cdots \lambda_{i_n} \Psi_{i_1 \ldots i_n} = \Psi(\lambda) ,\qquad
\lambda_{i_1} \cdots \lambda_{i_n}
\mathrm{T}_{i_1\ldots i_n}^{j_1 \ldots j_n}
 \mu_{j_1} \cdots \mu_{j_n} = \mathrm{T}(\lambda\,|\,\mu).
\end{gather}
Thus the symmetization over spinor indices is taken into account automatically.
Henceforth, in place of the tensors we work with the corresponding generating functions
which are homogeneous polynomials of degree $n$ of two variables
\begin{gather} \label{PsiHom}
\Psi(\lambda) = \Psi(\lambda_1,\lambda_2) , \qquad
\Psi(\alpha\lambda_1,\alpha\lambda_2) = \alpha^n \Psi(\lambda_1,\lambda_2).
\end{gather}
$\mathrm{T}(\lambda\,|\,\mu)$ is usually called the {\it symbol} of the operator.
In this way formula \eqref{TPsi} acquires a rather compact form
\begin{gather} \label{TPsispinor}
\left[\mathrm{T} \Psi\right](\lambda) =
\tfrac{1}{n!} \left.\mathrm{T}(\lambda\,|\,\partial_{\mu})
 \Psi(\mu)\right|_{\mu = 0}.
\end{gather}
Note that, in fact, we do not need to take $\mu = 0$ in \eqref{TPsispinor}.
The $\mu$ variable disappears automatically since $\mathrm{T}(\lambda\,|\,\mu)$
and $\Psi(\mu)$ have equal homogeneity degrees.

In order to illustrate the merits of auxiliary spinors
let us apply them to the text-book example of the quantum-mechanical system
of spin $\frac{n}{2}$,
i.e., consider the symmetry group $\mathrm{SU}(2)$ and the
generators $\vec{J}$ of the Lie algebra $su_2$
in the representation of spin $\frac{n}{2}$.
In the spin $\frac{1}{2}$ representation the generators act on the space $\mathbb{C}^2$ and
they are given by the Pauli matrices $\frac{\vec{\sigma}}{2}$, so that
\begin{gather*}
\big[\vec{J} \Psi\big]_i = \tfrac{1}{2} \vec{\sigma}_{i}^{j} \Psi_j
, \qquad
\vec{J}_{i}^{j} = \tfrac{1}{2} \vec{\sigma}_{i}^{j}.
\end{gather*}
Here the lower indices enumerate the rows and the upper indices -- the columns.
Taking the tensor product of $n$ spin $\frac{1}{2}$ representations
we obtain the generators on the space $\bigl(\mathbb{C}^2\bigr)^{\otimes n}$,
\begin{gather} \label{Jind}
\vec{J}_{i_1\ldots i_n}^{j_1 \ldots j_n} =
\tfrac{1}{2} \vec{\sigma}_{i_1}^{j_1}
\delta_{i_2}^{j_2}\cdots\delta_{i_n}^{j_n} + \cdots +
\tfrac{1}{2}
\delta_{i_1}^{j_1}\cdots\delta_{i_{n-1}}^{j_{n-1}}
\vec{\sigma}_{i_n}^{j_n}.
\end{gather}
In order to single out in the tensor product an irreducible maximal spin representation
we symmetrize over spinor indices yielding the representation of spin $\frac{n}{2}$,
\begin{gather} \label{JPsiind}
\big[\vec{J} \Psi\big]_{(i_1\ldots i_n)} = \tfrac{1}{2} \vec{\sigma}_{i_1}^{j} \Psi_{(j i_2 \ldots i_n)} +
\cdots + \tfrac{1}{2} \vec{\sigma}_{i_n}^{j} \Psi_{(i_1 \ldots i_{n-1} j)} .
\end{gather}

Further we introduce a pair of auxiliary spinors and f\/ind the symbol $\vec{J}(\lambda,\mu)$ of the opera\-tor~$\vec{J}$~\eqref{Jind}
converting formula \eqref{Jind} to
\begin{gather}\label{J}
\vec{J}(\lambda\,|\,\mu) = \lambda_{i_1} \cdots \lambda_{i_n}
\vec{J}_{i_1\ldots i_n}^{j_1 \ldots j_n}
 \mu_{j_1} \cdots \mu_{j_n} = \tfrac{n}{2} \langle\lambda\,|\,\mu\rangle^{n-1}
\langle\lambda\,|\,\vec{\sigma}\,|\,\mu\rangle, \\ \langle\lambda|
= \left(\lambda_1,\lambda_2\right),\qquad |\mu\rangle = \left(
\begin{matrix}
 \mu_1 \\ \mu_2\end{matrix}
\right),\nonumber
\end{gather}
where $\langle\lambda\,|\,\mu\rangle = \lambda_1\mu_1+\lambda_2\mu_2$ and
$\langle\lambda\,|\,\vec{\sigma}\,|\,\mu\rangle = \lambda_i\vec{\sigma}_{i}^{j}\mu_j $
are symbols of the identity operator and Pauli matrices, respectively.
In view of \eqref{TPsispinor}, \eqref{J}, formula \eqref{JPsiind} acquires the indexless form
\begin{gather*}
\big[\vec{J} \Psi\big](\lambda_1,\lambda_2) = \tfrac{1}{n!}\tfrac{n}{2} \langle\lambda\,|\,\partial_{\mu}\rangle^{n-1}
\langle\lambda\,|\,\vec{\sigma}\,|\,\partial_{\mu}\rangle \Psi(\mu)\big|_{\mu=0} .
\end{gather*}
Consequently, instead of tensors and f\/inite-dimensional operators we deal with their symbols and generating functions.
Note that due to the homogeneity of~$\Psi$~\eqref{PsiHom},
$\langle\lambda\,|\,\mu\rangle^n$ is a symbol of the identity
operator def\/ined on the tensor product of~$n$ spaces
\begin{gather*}
\left.\frac{1}{n!} \langle\lambda\,|\,\partial_{\mu}\rangle^{n} \Psi(\mu)\right|_{\mu=0} =
\left.\frac{1}{n!} \partial^{n}_{\alpha}
e^{\alpha\langle\lambda\,|\,\partial_{\mu}\rangle}
\Psi(\mu)\right|_{\mu=0 ,\alpha=0} =
\left.\frac{1}{n!}\partial^{n}_{\alpha}
\Psi(\alpha\lambda)\right|_{\alpha=0}
 = \Psi(\lambda).
\end{gather*}
Then taking into account that
\begin{gather*}
\tfrac{n}{2} \langle\lambda\,|\,\mu\rangle^{n-1}
\langle\lambda\,|\,\vec{\sigma}\,|\,\mu\rangle = \tfrac{1}{2}
 \langle\lambda\,|\,\vec{\sigma}\,|\,\partial_{\lambda}\rangle \langle\lambda\,|\,\mu\rangle^{n},
\end{gather*}
we obtain an alternative expression for $\vec{J}$,
\begin{gather}\label{J2}
\big[\vec{J} \Psi\big](\lambda_1,\lambda_2) =
\tfrac{1}{2}
 \langle\lambda\,|\,\vec{\sigma}\,|\,\partial_{\lambda}\rangle
 \tfrac{1}{n!} \langle\lambda\,|\,\partial_{\mu}\rangle^{n} \Psi(\mu)\big|_{\mu=0} =
\tfrac{1}{2}\langle\lambda\,|\,\vec{\sigma}\,|\,\partial_{\lambda}
\rangle \Psi(\lambda).
\end{gather}
Thus we have realized the Lie algebra generators $\vec{J}$
as dif\/ferential operators on the space of homogeneous polynomials of two variables
(forming a projective space)
\begin{gather*}
J_{\pm} = \tfrac{1}{2} \langle\lambda\,|\,\sigma_1\pm i\sigma_2\,|\,\partial_{\lambda}\rangle = \langle\lambda\,|\,\sigma_\pm\,|\,\partial_{\lambda}\rangle, \qquad
J_{3} = \tfrac{1}{2} \langle\lambda\,|\,\sigma_3\,|\,\partial_{\lambda}\rangle
\end{gather*}
or, more explicitly,
\begin{gather} \label{JSchw}
J_{+} = \lambda_1\partial_{\lambda_2}
 , \qquad J_{-} = \lambda_2\partial_{\lambda_1}
 , \qquad J_{3} = \tfrac{1}{2}
 (\lambda_1\partial_{\lambda_1} -
\lambda_2\partial_{\lambda_2} ).
\end{gather}
This realization of the generators is known as the Jordan--Schwinger representation.
We can choose the homogeneous function $(\lambda_1 + x \lambda_2)^n$ (see~\eqref{PsiHom}) as
a generating function of the $(n+1)$-dimensional representation with an auxiliary
parameter~$x$.

One can easily proceed from the projective space to the space of polynomials
of one complex variable. Indeed, due to the homogeneity
\begin{gather*}
\Psi(\lambda_1,\lambda_2) = \lambda_2^n
\Psi\big( \tfrac{\lambda_1}{\lambda_2} ,1\big) = \lambda_1^n
\Psi\big(1, \tfrac{\lambda_2}{\lambda_1}\big)
\end{gather*}
all information about $\Psi(\lambda_1,\lambda_2)$
is encoded in a function of the ratio $\frac{\lambda_1}{\lambda_2}$ alone.
In order to make contact with the holomorphic set of the ${\mathfrak{sl}}(2,\mathbb{C})$ generators \eqref{Egl2} we choose
$\lambda_1 = -z, \lambda_2=1$ and rewrite the generators~\eqref{JSchw} in terms of the
variable~$z$
\begin{gather} \label{Jgen}
J_{+} = z^2\partial - n z
 , \qquad J_{-} = -\partial
 , \qquad J_{3} = z\partial - \tfrac{n}{2}.
\end{gather}
Furthermore, the generating function of the Jordan--Schwinger representation turns into the generating function
of one variable~$(x-z)^n$
(cf.~\eqref{genfunsl2}, recall~\eqref{redsl2alg}).

Before proceeding to the fusion procedure for the $\mathrm{SL}(2,\mathbb{C})$ group,
we remind construction of the L-operator.
We build the L-operator with a f\/inite-dimensional local quantum space
starting from the Yang ${\rm R}$-matrix.
The latter acts on the tensor product of two spin-$\frac{1}{2}$ representations
\begin{gather} \label{Yang}
\mathrm{R}(u) = u+\tfrac{1}{2}\left({\hbox{{1}\kern-.25em\hbox{l}}}+\vec{\sigma}\otimes\vec{\sigma}\right)
= \left(%
\begin{matrix}
 u+\tfrac{1}{2} +
\tfrac{1}{2} \sigma_3 & \sigma_{-} \\
 \sigma_{+} & u+\tfrac{1}{2} -
\tfrac{1}{2} \sigma_3
\end{matrix}
\right).
\end{gather}
Following the recipe from \cite{KRS81,KS81} we form the product of the Yang ${\rm R}$-matrices
\begin{gather} \label{Rind}
\mathrm{R}_{(i_1\ldots i_n)}^{(j_1 \ldots j_n)}(u) = \operatorname{Sym} \mathrm{R}_{i_1}^{j_1}(u) \mathrm{R}_{i_2}^{j_2}(u-1)
 \cdots \mathrm{R}_{i_n}^{j_n}(u-n+1),
\end{gather}
where the indices refer to the f\/irst space in \eqref{Yang},
\begin{gather} \label{Rfusfact}
\mathrm{R}_i^j(u) = \big(u+\tfrac{1}{2}\big) \delta_{i}^{j}+
\tfrac{1}{2} \vec{\sigma}_{i}^{j} \vec{\sigma},
\end{gather}
and $\operatorname{Sym}$ implies symmetrization with respect to $(i_1\ldots i_n)$ and
$(j_1 \ldots j_n)$.
In such a way one obtains an operator acting on the space of symmetric rank $n$
tensors, i.e., on the space of spin~$\frac{n}{2}$ representation, and on the
two-dimensional auxiliary space where the $\vec{\sigma}$-matrices are acting.
According to~\cite{KRS81,KS81} it respects the Yang--Baxter relations.
Now we calculate the symbol of~\eqref{Rind} with respect to the quantum space
\begin{gather*}
\mathrm{R}(u\,|\,\lambda,\mu) = \lambda_{i_1} \cdots \lambda_{i_n}
\mathrm{R}_{i_1\ldots i_n}^{j_1 \ldots j_n}(u)
 \mu_{j_1} \cdots \mu_{j_n} \\
 \hphantom{\mathrm{R}(u\,|\,\lambda,\mu)}{}
 = \langle\lambda\,|\,\mathrm{R}(u)\,|\,\mu\rangle
\langle\lambda\,|\,\mathrm{R}(u-1)\,|\,\mu\rangle\cdots
\langle\lambda\,|\,\mathrm{R}(u-n+1)\,|\,\mu\rangle,
\end{gather*}
i.e., it is still an operator in the auxiliary space. Henceforth
for the sake of brevity we refer to it as a symbol of the ${\rm R}$-matrix.
The derived symbol $\mathrm{R}(u\,|\,\lambda,\mu)$ factorizes to a product of Yang's ${\rm R}$-matrix symbols
$\langle\lambda|\mathrm{R}(u)\,|\,\mu\rangle = \lambda_{i} \mathrm{R}_{i}^{j}(u) \mu_{j}$,
\begin{gather} \label{RYangSymb}
\langle\lambda\,|\,\mathrm{R}(u)\,|\,\mu\rangle = \langle\lambda\,|\,\mu\rangle
 \big(u \!+\!\tfrac{1}{2}\!+\!
\tfrac{1}{2} \vec{n} \vec{\sigma}\big)=
\left(
\begin{matrix}
 (u+1) \lambda_1\mu_1\!+\!u \lambda_2\mu_2\!\! & \lambda_2\mu_1 \\
 \lambda_1\mu_2 & \!\! u \lambda_1\mu_1\!+\! (u\!+\!1) \lambda_2\mu_2
\end{matrix}
\right),\!\!
\end{gather}
where we introduced the unit vector
$\vec{n} = \frac{\langle\lambda\,|\,\vec{\sigma}\,|\,\mu\rangle}{\langle\lambda\,|\,\mu\rangle} $,
$\vec{n}\cdot\vec{n} = 1 $.
The product of such matrices is easy to calculate and we obtain
\begin{gather}\label{matrix}
\mathrm{R}(u\,|\,\lambda,\mu) = u(u-1)\cdots(u-n+1)
\\
\times\!\left(
\begin{matrix}
 (u{+}1{-}\tfrac{n}{2})\langle\lambda\,|\,\mu\rangle^n{+}
 \tfrac{n}{2} \langle\lambda\,|\,\mu\rangle^{n{-}1}
 (\lambda_1\mu_1{-}\lambda_2\mu_2)\!\!\! & n\langle\lambda\,|\,\mu\rangle^{n-1}\lambda_2\mu_1 \\
 n\langle\lambda\,|\,\mu\rangle^{n-1}\lambda_1\mu_2
 &
 \!\!\!\! (u{+}1{-}\tfrac{n}{2})\langle\lambda\,|\,\mu\rangle^n{-}
 \tfrac{n}{2} \langle\lambda\,|\,\mu\rangle^{n{-}1}
 (\lambda_1\mu_1{-}\lambda_2\mu_2)
\end{matrix}
\right)\!.\nonumber
\end{gather}
In compact notation this formula takes the form
\begin{gather*}
\mathrm{R}(u\,|\,\lambda,\mu) = \langle\lambda\,|\,\mu\rangle^n
\big(u +\tfrac{1}{2} +
\tfrac{1}{2} \vec{n} \vec{\sigma}\big)
\big(u-\tfrac{1}{2}+
\tfrac{1}{2} \vec{n} \vec{\sigma}\big)
\cdots
\big(u-n+\tfrac{3}{2}+
\tfrac{1}{2} \vec{n} \vec{\sigma}\big)
\\
\hphantom{\mathrm{R}(u\,|\,\lambda,\mu)}{}
= u(u-1)\cdots(u-n+1)
 \langle\lambda\,|\,\mu\rangle^n \big(u+1-\tfrac{n}{2}+
\tfrac{n}{2} \vec{n} \vec{\sigma}\big)
\end{gather*}
and it can be easily proven by induction
using the identity $\left(\vec{n} \vec{\sigma}\right)^2 = {\hbox{{1}\kern-.25em\hbox{l}}} $.

Up to the inessential normalization factor
$r_n(u)=u(u-1)\cdots(u-n+1)$ and the shift
of the spectral parameter $u \to u-1+\frac{n}{2}$, we obtain the following symbol (see \eqref{J})
\begin{gather}
\mathrm{L}(u\,|\,\lambda,\mu) = r^{-1}_n(u) \mathrm{R}(u-1+\tfrac{n}{2}\,|\,\lambda,\mu) \nonumber\\
\hphantom{\mathrm{L}(u\,|\,\lambda,\mu)}{}
= u \langle\lambda\,|\,\mu\rangle^n + \tfrac{n}{2} \langle\lambda\,|\,\mu\rangle^{n-1}
\langle\lambda\,|\,\vec{\sigma}\,|\,\mu\rangle \vec{\sigma} = u \langle\lambda\,|\,\mu\rangle^n + \vec{J}(\lambda,\mu) \vec{\sigma}\label{symLLL}
\end{gather}
for the higher-spin ${\rm R}$-operator which acts on the tensor product
of the spin $\frac{n}{2}$ and spin $\frac{1}{2}$ representations.
Such an ${\rm R}$-operator is usually called the Lax operator with an $(n+1)$-dimensional local quantum space.
Let us emphasize once more that \eqref{symLLL} is a symbol of the Lax operator solely with respect
to the local quantum space, but it is a matrix in the $2$-dimensional auxiliary space.
In order to avoid misunderstandings we showed in~(\ref{matrix}) its explicit matrix form.
The expression $\langle\lambda\,|\,\mu\rangle^n$ is a symbol of the unit operator and
$\vec{J}(\lambda,\mu)$ is a symbol of the Lie algebra generators. Hence the fusion procedure yields
the familiar Lax operator,
\begin{gather} \label{LJ}
\mathrm{L}(u) = u {\hbox{{1}\kern-.25em\hbox{l}}} + \vec{J}\vec{\sigma} = \left(
\begin{matrix}
 u+J_3 & J_{-} \\
 J_{+} & u-J_3 \\
\end{matrix}
\right).
\end{gather}
The auxiliary spinors enabled us to reproduce this well-known result
in a remarkably simple and explicit way. They saved us from the need to
construct projectors which single out irreducible representations
and which are inevitable in the standard formulation.

Now we are going to describe another way
for deriving the L-operator~\eqref{LJ} by means of the fusion procedure.
The main reason to embark upon one more calculation is that
it can be generalized easily to the case of $q$-deformation (see Section~\ref{3.4})
and, more importantly, to the elliptic deformation~\cite{CDS}.
As before we deal with the symbols
of f\/inite-dimensional operators. The new ingredient is
a factorization of the ${\rm L}$-operator (cf.~\eqref{Lsl2}).
For calculating the symbol $\mathrm{R}(u\,|\,\lambda,\mu)$ of the ``fused'' ${\rm R}$-matrices~\eqref{Rind}
\begin{gather} \label{Yangstring}
\mathrm{R}(u\,|\,\lambda,\mu) = \langle\lambda\,|\,\mathrm{R}(u)\,|\,\mu\rangle
\langle\lambda\,|\,\mathrm{R}(u-1)\,|\,\mu\rangle\cdots
\langle\lambda\,|\,\mathrm{R}(u-n+1)\,|\,\mu\rangle,
\end{gather}
we choose the parametrization of the auxiliary spinor $\lambda_1 = -z$, $\lambda_2=1$
from the very beginning.
Remind a realization of the spin $\frac{1}{2}$ generators as dif\/ferential operators (cf.~\eqref{Jgen})
\begin{gather*}
J_{+} = z^2\partial - z
 , \qquad J_{-} = -\partial
 , \qquad J_{3} = z\partial -\tfrac{1}{2},
\end{gather*}
which act in the two-dimensional space of linear functions $\psi(z) = a_1 z+a_0$.
In the basis $\mathbf{e}_1 = -z$, $\mathbf{e}_2=1$ of this space
the matrices of the generators coincide with the Pauli-matrices
\begin{gather}
J_{\pm}\left(\mathbf{e}_1, \mathbf{e}_2\right) =
\left(J_{\pm}\mathbf{e}_1, J_{\pm}\mathbf{e}_2\right)=
\left(\mathbf{e}_1, \mathbf{e}_2\right) \sigma_{\pm}, \nonumber\\
J_{3}\left(\mathbf{e}_1, \mathbf{e}_2\right) =
\left(J_{3}\mathbf{e}_1, J_{3}\mathbf{e}_2\right) =
\left(\mathbf{e}_1, \mathbf{e}_2\right) \tfrac{1}{2}\sigma_{3}.\label{basis}
\end{gather}
Next we use the fusion procedure and derive the Lax operator \eqref{LJ} together with
a representation of the spin $\frac{n}{2}$ generators \eqref{Jgen}
acting in the $(n+1)$-dimensional space of
polynomials $\psi(z) = a_n z^n+\cdots+a_0$.

The symbol $\langle\lambda\,|\,\mathrm{R}(u)\,|\,\mu\rangle$ of Yang's ${\rm R}$-matrix
has been already found above \eqref{RYangSymb}, but now we are going to rewrite it in a dif\/ferent form.
We represent it as a dif\/ferential operator in the spinor variables
acting on the identity operator symbol.
Indeed, let us rewrite relations~(\ref{basis}) in the equivalent form
\begin{gather}\label{basis1}
(-z, 1) \sigma_{\pm} = J_{\pm}(-z, 1), \qquad
(-z, 1) \tfrac{1}{2}\sigma_{3} =
J_{3}(-z, 1),
\end{gather}
and use these formulae for calculating the symbol of Yang's ${\rm R}$-matrix \eqref{Yang}
\begin{gather*}
\langle\lambda\,|\,\mathrm{R}(u)\,|\,\mu\rangle
= \left(
\begin{matrix}
 (-z, 1)
 \left(u+\tfrac{1}{2} +
\tfrac{1}{2} \sigma_3\right)|\, \mu\rangle &
(-z, 1)\sigma_{-}\,|\,\mu\rangle \\
 (-z, 1)\sigma_{+}\,|\,\mu\rangle & (-z, 1)\left(u+\tfrac{1}{2} -
\tfrac{1}{2} \sigma_3\right)|\,\mu\rangle
\end{matrix}
\right)\\
\hphantom{\langle\lambda\,|\,\mathrm{R}(u)\,|\,\mu\rangle}{} = \left(
\begin{matrix}
 u+z\partial & -\partial \\
 z^2\partial-z & u+1-z\partial
\end{matrix}
\right)\left(\mu_2-\mu_1 z\right).
\end{gather*}
We obtained the spin $\ell = \frac{1}{2}$ L-operator \eqref{LaxNonFact}
(with the shifted spectral parameter $u \to u + \frac{1}{2}$) acting on
the symbol of the identity operator
$\langle \lambda \,|\, \mu\rangle = \left(\mu_2-\mu_1 z\right)$.
Then we observe that this symbol can be cast in the factorized form
\begin{gather} \label{LnewFact}
\langle\lambda\,|\,\mathrm{R}(u)\,|\,\mu\rangle =\begin{pmatrix}
 1 & 0 \\ z & u+1
 \end{pmatrix}
 \begin{pmatrix}
 1 & -\partial_1 \\ 0 & 1
 \end{pmatrix}
 \begin{pmatrix}
 u & 0 \\ -z & 1
 \end{pmatrix}\left.
 \left(\mu_2-\mu_1 z_1\right)\right|_{z_1=z},
\end{gather}
which is easily checked by a direct calculation. Note that the factorization
in \eqref{LnewFact} is slightly dif\/ferent from \eqref{Lsl2} (at $\ell = \frac{1}{2}$),
since it involves a particular ordering of $z$ and $\partial$
and such ordering is compatible with the factorization of
the L-operator up to the shift of spectral parameter.

Then we consider the product of two consecutive symbols in \eqref{Yangstring}
and prof\/it a lot from the factorization \eqref{LnewFact} which provides cancellation of two
adjacent matrix factors (which are underlined in the following formula)
\begin{gather*}
\langle\lambda\,|\,\mathrm{R}(u)\,|\,\mu\rangle
\langle\lambda\,|\,\mathrm{R}(u-1)\,|\,\mu\rangle
=\begin{pmatrix}
1 & 0 \\ z & u+1
\end{pmatrix}
\begin{pmatrix}
1 & -\partial_1 \\ 0 & 1
\end{pmatrix}
\underline{\begin{pmatrix}
u & 0 \\ -z & 1
\end{pmatrix}\begin{pmatrix}
1 & 0 \\ z & u
\end{pmatrix}}
\\
\qquad\quad{}\times\begin{pmatrix}
1 & -\partial_2 \\ 0 & 1
\end{pmatrix}
\begin{pmatrix}
u-1 & 0 \\ -z & 1
\end{pmatrix}
\left.
\left(\mu_2-\mu_1 z_1\right)
\left(\mu_2-\mu_1 z_2\right)
\right|_{z_1=z_2=z}
\\
\qquad{}
= u \begin{pmatrix}
1 & 0 \\ z & u+1
\end{pmatrix}
\begin{pmatrix}
1 & -\partial_1-\partial_2 \\ 0 & 1
\end{pmatrix}
\begin{pmatrix}
u-1 & 0 \\ -z & 1
\end{pmatrix}
\left.
\left(\mu_2-\mu_1 z_1\right)
\left(\mu_2-\mu_1 z_2\right)
\right|_{z_1=z_2=z}.
\end{gather*}
By now the generalization of the previous result to the product of $n-1$ symbols~\eqref{Yangstring} is evident
\begin{gather*}
\langle\lambda\,|\,\mathrm{R}(u)\,|\,\mu\rangle
\langle\lambda\,|\,\mathrm{R}(u-1)\,|\,\mu\rangle\cdots
\langle\lambda\,|\,\mathrm{R}(u-n+1)\,|\,\mu\rangle
\\
\qquad{}
= r_n(u)
\begin{pmatrix}
1 & 0 \\ z & u+1
\end{pmatrix}
\begin{pmatrix}
1 & -\partial_1-\partial_2-\cdots-\partial_n \\ 0 & 1
\end{pmatrix}
\begin{pmatrix}
u-n+1 & 0 \\ -z & 1
\end{pmatrix}\\
\qquad\quad{}\times
\left.
\left(\mu_2-\mu_1 z_1\right)\cdots
\left(\mu_2-\mu_1 z_n\right)
\right|_{z_1=\cdots=z_n=z}.
\end{gather*}
Further we multiply all matrices on the right-hand side of the previous formula and obtain
\begin{gather*}
r_n(u) \begin{pmatrix}
u-n+1 +z(\partial_1+\cdots+\partial_n)& -\partial_1-\cdots-\partial_n \\
z^2(\partial_1+\cdots+\partial_n)-n z & u+1 -z(\partial_1+\cdots+\partial_n)
\end{pmatrix}
\\
\qquad\quad{}\times
\left.
\left(\mu_2-\mu_1 z_1\right)\cdots
\left(\mu_2-\mu_1 z_n\right)
\right|_{z_1=\cdots=z_n=z}\\
\qquad{} =
r_n(u) \begin{pmatrix}
u-n+1 +z\partial& -\partial \\
z^2\partial-n z & u+1 -z\partial
\end{pmatrix}
\left(\mu_2-\mu_1 z\right)^n,
\end{gather*}
where on the last step we use an obvious formula
\begin{gather*}
(\partial_1+\cdots+\partial_n)\left.
\left(\mu_2-\mu_1 z_1\right)\cdots
\left(\mu_2-\mu_1 z_n\right)
\right|_{z_1=\cdots=z_n=z} = \partial \left(\mu_2-\mu_1 z\right)^n.
\end{gather*}
The f\/inal result for the symbol \eqref{Yangstring} of the ``fused'' Yang ${\rm R}$-matrices is
\begin{gather*}
\mathrm{R}(u\,|\,\lambda,\mu) = r_n(u) \begin{pmatrix}
u+1-\tfrac{n}{2}+J_3& J_- \\
J_+ & u+1-\tfrac{n}{2}-J_3
\end{pmatrix}
\left(\mu_2-\mu_1 z\right)^n,
\end{gather*}
where the generators $J_{\pm}$, $J_3$ for the representation of spin $\frac{n}{2}$
are given by~(\ref{Jgen}).

Factorization of the ${\rm L}$-operator plays an important role in the
construction of the general ${\rm R}$-operator for deformed \cite{Derkachov:2007gr,DS1}
and non-deformed \cite{Derkachov:2007gr} rank 1 symmetry algebra,
as well as in the higher rank case \cite{DM09}.
Here we see that it f\/inds a natural place
in the fusion construction as well.

\subsection[Fusion construction for $\mathrm{SL}(2,\mathbb{C})$]{Fusion construction for $\boldsymbol{\mathrm{SL}(2,\mathbb{C})}$}
\label{fusSL2}

The fusion procedure enables one to produce even more intricate ${\mathfrak{sl}}_2$-symmetric solutions of the
YBE. Starting from the L-operator acting in the tensor product of
spin~$\frac{1}{2}$ and spin $\ell$ representations (we assume~$\ell$ to be generic
such that the corresponding representation is inf\/inite-dimensional)
one obtains the ${\rm R}$-operator which acts in the tensor product of
spin~$\frac{n}{2}$ and spin~$\ell$ representations.
Since we are mainly interested in the ${\rm R}$-operators
which are invariant with respect to the~$\mathrm{SL}(2,\mathbb{C})$ group,
we will thoroughly study how the fusion procedure applies in this case.

As before we prof\/it a lot from the auxiliary spinors notation.
However, from now on the holomorphic and antiholomorphic sectors are present
and we need to introduce a pair of auxiliary spinors $\lambda_{i}$, $\bar{\lambda}_{\bar{i}}$.
They are independent variables not related by the complex conjugation.
Thus we introduce a pair of scalar objects (without spinor indices)
\begin{gather}
\Lambda(u,\lambda,\mu) = \lambda_{i} \mathrm{L}^{j}_{i}(u) \mu_{j} ,\qquad
 \bar{\Lambda}(\bar{u},\bar{\lambda},\bar{\mu}) =
\bar{\lambda}_{\bar{i}} \bar{\mathrm{L}}^{\bar{j}}_{\bar{i}}(\bar{u}) \bar{\mu}_{\bar{j}},
\end{gather}
which are linear combinations of the ${\rm L}$-operators' entries \eqref{Lsl2}, \eqref{barLsl2}.
An easy calculation shows that\footnote{We should note that the idea to reformulate the
fusion procedure with the help of such $\Lambda$-operators belongs to D.~Karakhanyan.}
\begin{gather}
\Lambda(u,\lambda,\mu) = - (\lambda_1 + \lambda_2 z)(\mu_2 - \mu_1 z)\partial +u_2 \lambda_2 (\mu_2 - \mu_1 z)-(u_1+1) \mu_1 (\lambda_1 + \lambda_2 z)
\nonumber\\ \label{Lam}
\hphantom{\Lambda(u,\lambda,\mu)}{}
= - (\lambda_1 + \lambda_2 z)^{u_2+1} (\mu_2 - \mu_1 z)^{-u_1}
\cdot\partial\cdot
(\lambda_1 + \lambda_2 z)^{-u_2} (\mu_2 - \mu_1 z)^{u_1 + 1}.
\end{gather}
The factorized expression \eqref{Lam} looks much like formula \eqref{Lsl2}.
Indeed in both expressions the dif\/ferential operators are sandwiched
between some multiplication by a~function operators.
The analogous relation takes place for~$\bar{\mathrm{L}}$~\eqref{barLsl2}.
Then we multiply a number of $\Lambda$-operators with shifted spectral
parameters to form a $\Lambda$-string
\begin{gather*}
\Lambda(u) \Lambda(u-1) \cdots \Lambda(u - n + 1) =
 (-1)^n (\lambda_1 + \lambda_2 z)^{u_2+1} (\mu_2 - \mu_1 z)^{-u_1}
 \\
 \qquad\quad{}\times
 \left(\partial (\mu_2 - \mu_1 z)^2\right)^n\cdot
(\lambda_1 + \lambda_2 z)^{-u_2+n-1} (\mu_2 - \mu_1 z)^{u_1 -n} =
\\
\qquad{}
= (-1)^n (\lambda_1 + \lambda_2 z)^{u_2+1} (\mu_2 - \mu_1 z)^{-u_1+n-1}
\cdot\partial^n\cdot
(\lambda_1 + \lambda_2 z)^{-u_2+n-1} (\mu_2 - \mu_1 z)^{u_1+1}.
\end{gather*}
Here we apply the formula which can be easily proven by induction,
\begin{gather*}
\left(\partial (\mu_2 - \mu_1 z)^2\right)^n = (\mu_2 - \mu_1 z)^{n-1}
 \partial^n (\mu_2 - \mu_1 z)^{n+1}.
\end{gather*}

Then we take into account the anti-holomorphic sector and form the product
of $\Lambda$- and $\bar\Lambda$-strings resulting in the symbol for a higher-spin
${\rm R}$-operator
\begin{gather}
\mathrm{R}_{\text{fus}}(u,\bar{u}\,|\,\lambda,\bar{\lambda},\mu,\bar{\mu})
= \Lambda(u) \Lambda(u-1) \cdots \Lambda(u - n + 1)
\bar{\Lambda}(\bar{u}) \bar{\Lambda}(\bar{u}-1) \cdots \bar{\Lambda}(\bar{u} - \bar{n} + 1) \label{Lamline}\\ =
(-1)^{n+\bar n} [\lambda_1 + \lambda_2 z]^{u_2 + 1}
[\mu_2 - \mu_1 z]^{- u_1 + n - 1} \cdot [\partial_z]^n \cdot
[\lambda_1 + \lambda_2 z]^{- u_2 + n - 1}
[\mu_2 - \mu_1 z]^{u_1 + 1}.\notag
\end{gather}
This ${\rm R}$-operator acts in the tensor product of the inf\/inite-dimensional representation
specif\/ied by the spins~$\ell$,~$\bar{\ell}$
and the f\/inite-dimensional representation with the spins $\frac{n}{2}$, $\frac{\bar{n}}{2}$.
Let us remind that~$\mathrm{R}_{\text{fus}}$ is a symbol with respect to the
f\/irst f\/inite-dimensional space only,
but it is a dif\/ferential operator in the second inf\/inite-dimensional space.
Evidently, the right-hand side of~\eqref{Lamline} is polynomial in~$\lambda$ and~$\mu$, as it should be.
In order to reconstruct the operator itself from its symbol we resort to the rule~\eqref{TPsispinor}.
More precisely, we apply the corresponding relation to a~function
$\Phi(\lambda,\bar{\lambda}\,|\,z,\bar{z})$, which is homogeneous in $\lambda$ and
$\bar{\lambda}$ of the homogeneity degree~$n$ and~$\bar{n}$, respectively,
\begin{gather} \label{fussl2}
\left[ \mathrm{R}_{\text{fus}}(u,\bar{u}) \Phi \right](\lambda,\bar{\lambda}\,|\,z,\bar{z})
= \left. \mathrm{R}_{\text{fus}}(u,\bar{u}\,|\,\lambda,\bar{\lambda},\partial_{\mu},\partial_{\bar{\mu}})
\Phi(\mu,\bar{\mu}\,|\,z,\bar{z}) \right|_{\mu = \bar{\mu}= 0}.
\end{gather}
We stress that the fusion formulae \eqref{Lamline}, \eqref{fussl2}
are somewhat dif\/ferent from the standard ones. We f\/ind them better adapted
for applications.

The higher-spin ${\rm R}$-operator \eqref{fussl2} obtained by means of the fusion is
identical with the reduction of the general ${\rm R}$-operator calculated in \eqref{redsl2},
which will be demonstrated shortly. First of all the operator form of the star-triangle relation \eqref{unique}
enables one to rewrite the symbol~\eqref{Lamline} as follows
\begin{gather}
\mathrm{R}_{\text{fus}}(u,\bar{u}\,|\,\lambda,\bar{\lambda},\mu,\bar{\mu}) =
(-1)^{n+\bar n} [\lambda_1 + \lambda_2 z]^{u_2 + 1}\nonumber\\
\hphantom{\mathrm{R}_{\text{fus}}(u,\bar{u}\,|\,\lambda,\bar{\lambda},\mu,\bar{\mu}) =}{}\times
[\partial_z]^{u_1 + 1} \cdot [\mu_2 - \mu_1 z]^n \cdot
[\partial_z]^{- u_1 + n - 1}
[\lambda_1 + \lambda_2 z]^{- u_2 + n - 1}. \label{Lamline2}
\end{gather}
We have seen above that $(\lambda_1 + \lambda_2 x)^n$ is a generating function of the $(n+1)$-dimensional
Jordan--Schwinger representation of~${\mathfrak{sl}}_2$. Its generalization to the group
$\mathrm{SL}(2,\mathbb{C})$ is straightforward:
$[\lambda_1 + \lambda_2 x]^n$ (see~\eqref{not1}) is a generating function of
the $(n+1)(\bar{n}+1)$-dimensional representation realized in the space
of homogeneous functions~$\Psi(\lambda,\bar{\lambda})$,
\begin{gather*}
\Psi(\lambda,\bar{\lambda}) = \Psi(\lambda_1,\lambda_2,\bar{\lambda}_1,\bar{\lambda}_2) , \qquad
\Psi(\alpha\lambda_1,\alpha\lambda_2,\bar{\alpha}\bar{\lambda}_1,\bar{\alpha}\bar{\lambda}_2)
= \alpha^{n} \bar{\alpha}^{\bar{n}} \Psi(\lambda,\bar{\lambda}).
\end{gather*}
Then we act by $\mathrm{R}_{\text{fus}}$ on the generating function according to~\eqref{fussl2}
and choose the symbol in the form~\eqref{Lamline2}. At the same time we do not act
by the $\mathrm{R}_{\text{fus}}$-operator on any function in its second space. At this point we
take into account that
\begin{gather}
[\partial_{\mu_2} - \partial_{\mu_1} z]^n [\mu_1 + \mu_2 x]^n = n!\bar{n}! [x - z]^n
\end{gather}
and obtain
\begin{gather}
\mathrm{R}_{\text{fus}}\big(u+\tfrac{n}{2},\bar{u}+\tfrac{\bar{n}}{2}\big) [\lambda_1 + \lambda_2 x]^n
 \notag
\\ \qquad{}
= n!\bar{n}! [\lambda_1 + \lambda_2 z]^{u + \frac{n}{2} +\ell + 1} [\partial_z]^{u + \frac{n}{2} - \ell} [z - x]^n
 [\partial_z]^{- u + \frac{n}{2} + \ell}
 [\lambda_1 + \lambda_2 z]^{- u + \frac{n}{2} - \ell - 1} \notag
\\ \qquad{}
= n!\bar{n}! [\lambda_1+\lambda_2 z]^{u + \frac{n}{2} +\ell + 1} [z - x]^{- u + \frac{n}{2} + \ell}
 [\partial_z]^n [z - x]^{u + \frac{n}{2} - \ell}
 [\lambda_1+\lambda_2 z]^{- u + \frac{n}{2} - \ell - 1}.
\label{fusgensl2} \end{gather}
Here we prof\/ited from the star-triangle relation \eqref{unique} at the last step.
In order to compare the reduction of the general ${\rm R}$-operator \eqref{redsl2}
with the expression \eqref{fusgensl2} following from the fusion formula \eqref{Lamline}
we just need to pass from the Jordan--Schwinger
representation to the standard representation of $\mathrm{SL}(2,\mathbb{C})$ (in
the space of functions of one complex variable)
described in Section~\ref{SL2rep}. Consequently we choose
$\lambda_1 = - z_1$,
$\bar{\lambda}_1 = - \bar{z}_1$, $\lambda_2 = \bar{\lambda}_2 = 1$
and denote $z = z_2$, $\bar{z} = \bar{z}_2$.
Finally, we see that both formulae are identical up to a numerical normalization.

We conclude that both ways to construct the higher-spin
f\/inite-dimensional (in one of the spaces) ${\rm R}$-operators give identical results,
and the general ${\rm R}$-operator~\eqref{R} contains all solutions of the Yang--Baxter
equation associated with
the principal series representations of the $\mathrm{SL}(2,\mathbb{C})$ group.

\section{The Faddeev modular double}

Using the patterns of the previous sections, in the following we show that all described
constructions for the group $\mathrm{SL}(2,\mathbb{C})$ can be straightforwardly adapted
to the modular double. Additionally, we construct corresponding f\/inite-dimensional
solutions of the YBE using the fusion.

\subsection{Representations of the quantum algebra}
\label{qqsl2}

The modular double of $\mathcal{U}_q({\mathfrak{sl}}_2)$ was introduced by Faddeev in~\cite{F99}.
This algebra is formed by two sets of generators
$\mathbf{E}$, $\mathbf{F}$, $\mathbf{K}$ and
$\widetilde{\mathbf{E}}$, $\widetilde{\mathbf{F}}$, $\widetilde{\mathbf{K}}$.
The usual commutation relations for $\mathbf{E}$, $\mathbf{F}$, $\mathbf{K}$ which generate~$\mathcal{U}_q({\mathfrak{sl}}_2)$
with $q = e^{i \pi \tau}$ ($\tau \in \mathbb{C}$ and it is not a rational number)
\begin{gather} \label{qsl2}
[\mathbf{E},\mathbf{F}] = \frac{\mathbf{K}^2 - \mathbf{K}^{-2}}{q-q^{-1}} ,\qquad
\mathbf{K} \mathbf{E} = q \mathbf{E} \mathbf{K} ,\qquad
\mathbf{K} \mathbf{F} = q^{-1} \mathbf{F} \mathbf{K},
\end{gather}
are supplemented by similar relations for $\widetilde{\mathbf{E}}$, $\widetilde{\mathbf{F}}$, $\widetilde{\mathbf{K}}$
with the deformation parameter $\widetilde{q} = e^{i \pi / \tau}$. The generators
$\mathbf{E}$ and $\mathbf{F}$ commute with $\widetilde{\mathbf{E}}$
and $\widetilde{\mathbf{F}}$.
The generator $\mathbf{K}$ anti-commutes with~$\widetilde{\mathbf{E}}$ and~$\widetilde{\mathbf{F}}$,~$\widetilde{\mathbf{K}}$ anti-commutes with~$\mathbf{E}$ and $\mathbf{F}$ while $\mathbf{K}$ commutes with~$\widetilde{\mathbf{K}}$.

For particular representations of the modular double
see~\cite{BT02,F99,FKV,PT, BultPhD} and references therein.
We use the parametrization $\tau = \frac{\omega'}{\omega}$, where
$\omega$ and~$\omega'$ are complex numbers with the positive imaginary parts,
$\operatorname{Im} \omega >0$, $\operatorname{Im} \omega' >0$, satisfying
the normalization condition $\omega \omega' = -\frac{1}{4}$. Then
\begin{gather*}
q = \exp\left(i \pi \omega' / \omega \right) ,\qquad
\widetilde{q} = \exp\left(i \pi \omega / \omega' \right),
\end{gather*}
and the change $q \rightleftarrows \widetilde{q}$ is equivalent to
$\omega \rightleftarrows \omega'$. We denote also
\begin{gather} \label{nota}
\omega'' = \omega + \omega'
,\qquad \beta = \frac{\pi}{12}\left( \frac{\omega}{\omega'} +
\frac{\omega'}{\omega} \right).
\end{gather}

In the following we deal with a representation $\pi_s$ of the modular double
when the generators $\mathbf{K}_s = \pi_s(\mathbf{K})$,
$\mathbf{E}_s = \pi_s(\mathbf{E})$, $\mathbf{F}_s = \pi_s(\mathbf{F})$
are realized as f\/inite-dif\/ference operators
acting on the space of entire functions rapidly decaying at inf\/inity
along contours parallel to the real line.
This representation is parameterized by one complex parameter~$s$ called the {\it spin},
and the generators have the following explicit form~\cite{BT02,BT06,CD14}
\begin{gather} \label{Gs}
\mathbf{K}_s = e^{-\frac{i \pi}{2\omega} \hat p} ,\qquad
\big(q-q^{-1}\big)\mathbf{E}_s =
e^{\frac{i \pi x}{\omega}} \Big[
e^{-\frac{i \pi}{2 \omega}\left(\hat p -s - \omega''\right)} -
e^{\frac{i \pi}{2 \omega}\left(\hat p -s - \omega''\right)}
\Big] ,\\
\big(q-q^{-1}\big)\mathbf{F}_s =
e^{-\frac{i \pi x}{\omega}} \Big[
e^{\frac{i \pi}{2 \omega}\left(\hat p + s + \omega''\right)} -
e^{-\frac{i \pi}{2 \omega}\left(\hat p + s + \omega''\right)}
\Big], \nonumber
\end{gather}
where $\hat p$ denotes a momentum operator in the coordinate
representation $\hat p = \frac{1}{2 \pi i} \partial_{x}$.
The formulae for generators
$\widetilde{\mathbf{K}}_s$,
$\widetilde{\mathbf{E}}_s$,
$\widetilde{\mathbf{F}}_s$ are
obtained by a simple interchange $\omega \rightleftarrows \omega'$ in~\eqref{Gs}.

The modular double is associated with two basic special functions.
The f\/irst one is the non-compact quantum dilogarithm
which has the following integral representation
\begin{gather} \label{gamma}
\gamma(z) = \exp\left(-\frac{1}{4}\int\limits^{+\infty}_{-\infty}
\frac{d t}{t} \frac{e^{i t z}}{\sin(\omega t)\sin(\omega^{\prime} t)}\right),
\end{gather}
where the contour goes above the singularity at $t = 0$.
In the context of quantum integrable systems it has been found f\/irst in~\cite{F95}.
Some basic formulae for $\gamma(z)$ can be found in~\cite{FKV,Vol05}.
This function respects a pair of f\/inite-dif\/ference equations of the f\/irst order
and the ref\/lection relation
\begin{gather} \label{gFunEq}
\frac{\gamma(z+\omega^{\prime})}
{\gamma(z-\omega^{\prime})} = 1+e^{-\frac{i\pi}{\omega} z} , \qquad
\frac{\gamma(z+\omega)}{\gamma(z-\omega)} = 1+e^{-\frac{i\pi}{\omega^{\prime}} z}, \qquad
\gamma(z) \gamma(-z) = e^{i\beta} e^{i\pi z^2}.
\end{gather}
One can interpret $2\omega$ and $2\omega'$ as some quasiperiods of
the quantum dilogarithm.

The second function we need is
\begin{gather} \label{D}
D_{a}(z) = e^{-2 \pi i a z} \frac{\gamma(z+a)}{\gamma(z-a)}.
\end{gather}
In fact it coincides with the Faddeev--Volkov ${\rm R}$-matrix~\cite{Bazhanov:2007mh, VF}.
Some relations for this functions are presented in~\cite{BT06}.
It naturally arises when one looks for the intertwining operator of equivalent
representations of the modular double~\cite{PT}, and it serves as the main building
block in the construction of a general ${\rm R}$-matrix as an integral operator
\cite{BT06,CD14}.
This general ${\rm R}$-operator is a product of four Faddeev--Volkov's ${\rm R}$-matrices.
The function $D_{a}(z)$ obeys simple ref\/lection relations
\begin{gather} \label{Dev}
D_a(z) = D_a(-z), \qquad D_a(z) D_{-a}(z) = 1,
\end{gather}
and a pair of f\/inite-dif\/ference equations of the f\/irst order
\begin{gather} \label{FunEq}
\frac{D_a(z-\omega')}{D_a(z + \omega')} =
\frac{\cos \frac{\pi}{2 \omega} ( z-a)}{\cos \frac{\pi}{2 \omega} ( z+a)}, \qquad
\frac{D_a(z-\omega)}{D_a(z + \omega)} =
\frac{\cos \frac{\pi}{2 \omega'} ( z-a)}{\cos \frac{\pi}{2 \omega'} ( z+a)}.
\end{gather}
Note that the functions $\gamma(z)$ and $D_{a}(z)$ are symmetric with respect
to $\omega$ and $\omega'$.

A generalization of the Faddeev--Volkov model still associated with the
$\gamma(z)$-function was found in \cite{spi:conm}.
It leads to a more general ${\rm R}$-operator than we consider here \cite{CS},
which can be obtained as a limit from the most
complicated known ${\rm R}$-operator derived in \cite{DS1}.

Now we proceed to f\/inite-dimensional representations of the modular double.
In order to f\/ix the spin $s_0$ such that a f\/inite-dimensional representation decouples
from the inf\/inite-dimensional representation $\pi_{s_0}$
we resort to the {\it intertwining} operator of equivalent
representations of the modular double.
It is known that the representations $\pi_s$ and $\pi_{-s}$ are equivalent.
The corresponding intertwining operator~\cite{PT}
is expressed in terms of the special function \eqref{D}, such that
\begin{gather} \label{intw1}
D_{-s}(\hat p) \mathbf{K}_s = \mathbf{K}_{-s} D_{-s}(\hat p),\!\qquad
D_{-s}(\hat p) \mathbf{E}_s = \mathbf{E}_{-s} D_{-s}(\hat p),\!\qquad
D_{-s}(\hat p) \mathbf{F}_s = \mathbf{F}_{-s} D_{-s}(\hat p),\!\!
\end{gather}
where $\hat p$ is the momentum operator.
There are analogous relations for $\widetilde{\mathbf{E}}$,
$\widetilde{\mathbf{F}}$, $\widetilde{\mathbf{K}}$, since
the $D$-function is invariant with respect to the permutation of~$\omega$ and~$\omega'$.
The latter relations can be easily checked using equations~\eqref{FunEq}.
Applying the Fourier transformation of the $D$-func\-tion~\eqref{D}~\cite{BT06,FKV,Vol05}
\begin{gather} \label{FourierD}
A(a) \int\limits^{+\infty}_{-\infty} d t \, e^{2 \pi i t z } D_{a}(t) =
D_{-\omega''-a}(z), \\
A(a) \equiv \frac{e^{\frac{i \pi}{2}(2a+\omega'')^2+\frac{i \beta}{2}}}{\gamma(2a+\omega'')}, \qquad
A(a)A(-a-\omega'') = 1,\nonumber
\end{gather}
we immediately represent the intertwiner $D_{-s}(\hat p)$ as an integral operator (in analogy with \eqref{dint})
\begin{gather} \label{Wint}
D_{-s}(\hat p) \Phi(x) =
A(s-\omega'') \int\limits^{+\infty}_{-\infty}d x' \, D_{s-\omega''}(x-x') \Phi(x').
\end{gather}
Thus the intertwiner admits two forms for generic values of $s$:
as a formal function of the momentum operator and a well-def\/ined integral operator.

A f\/inite $(n+1)(m+1)$-dimensional representation decouples from the
inf\/inite-dimensional one for special values of the spin
\begin{gather*}
s=s_0: = - \omega''- n \omega - m \omega', \qquad n , m \in \mathbb{Z}_{\geq 0},
\end{gather*}
where the integers $n$ and $m$ enumerate the points of a quarter-inf\/inite
lattice on the complex plane (or a line, for real~$\omega/\omega'$).
Such f\/inite-dimensional representations emerged f\/irst in the two-dimensional
rational conformal f\/ield theory through $6j$-symbols for the modular double with
$q=e^{2\pi im/(m+1)}$ and $\tilde q=e^{2\pi i(m+1)/m}$,
$m\in\mathbb{Z}$~\cite{2dcft}. The most general two-index $6j$-symbols were discovered
in the theory of elliptic hypergeometric functions~\cite{spi:umn2}.
They are described by meromorphic functions of one variable satisfying the biorthogonality
relation with an absolutely continuous measure, which has the form usually
ascribed to the functions of two independent variables. The latter property brought
to the theory of spectral problems the new notion of two-index (bi-)orthogonality.

For the elliptic modular double the two-index f\/inite-dimensional representations
were discovered in~\cite{DS1,DS2}. In principle, f\/inite-dimensional
representations of the Faddeev modular double can
be derived as certain limits from this elliptic construction,
but we give here an independent consideration and, moreover,
describe the f\/inite-dimensional ${\rm R}$-matrices analogous to~\eqref{redsl2alg}.
Note that in the $\mathrm{SL}(2,\mathbb{C})$ group case the f\/inite-dimensional
representations were also parametrized by a pair of non-negative integer numbers
$n$ and $\bar n$, but the integer $\bar n$ has a dif\/ferent nature emerging from
a discretization of the separate spin variable $\bar s$, which is absent in our case.

In order to f\/ind f\/inite-dimensional representations of interest we investigate
the null-space of the intertwiner. We take the formal operator identity
\begin{gather}
D_{-s}(\hat p) D_s(\hat p) = {\hbox{{1}\kern-.25em\hbox{l}}},
\label{form_inv}\end{gather}
which is a consequence of the ref\/lection formula \eqref{Dev},
and rewrite it in an equivalent form substituting $D_s(\hat p)$ and ${\hbox{{1}\kern-.25em\hbox{l}}}$ for their kernels (see \eqref{Wint})
\begin{gather} \label{DDdel}
D_{-s}(\hat p) D_{-s-\omega''}(x-y) = A^{-1}(-s-\omega'') \delta(x-y).
\end{gather}
Then we note that zeros of the quantum dilogarithm $\gamma(z) = 0$ are located
at $z = \omega'' + 2n \omega + 2 m \omega'$, $n,m \in \mathbb{Z}_{\geq 0} $,
which indicates that the relation \eqref{form_inv} is broken down at the
corresponding points.
Consequently for the spin values specif\/ied above, $s = s_0$, the right-hand
side of \eqref{DDdel} vanishes and a nontrivial null-space of $D_{-s_0}(\hat p)$ arises
\begin{gather*}
D_{-s_0}(\hat p) D_{n \omega + m \omega'}(x-y) = 0,\qquad n , m \in \mathbb{Z}_{\geq 0}.
\end{gather*}
The latter formula is a deformed analogue of \eqref{Wgenfunsl2}.
From the intertwining relations \eqref{intw1} the null-space is seen to be
invariant under the action of the modular double generators.
Corresponding representation is f\/inite-dimensional as we will see shortly.
The generators are f\/ixed by expressions \eqref{Gs} and their modular duals with the
spin parameter $s=s_0$.
One can also show that the corresponding representation is irreducible.

In this way we have found that $D_{n\omega+m\omega'}(x-y)$ is the
generating function of a f\/inite-dimensional representation
containing all its basis vectors. Here $y$ is an auxiliary parameter,
which is convenient to write in the exponential form
\begin{gather} \label{Yvar}
Y= Y(y) = e^{\frac{i \pi}{2\omega}y} ,\qquad
\widetilde{Y} = \widetilde{Y}(y) = e^{\frac{i \pi}{2\omega'}y}.
\end{gather}
Since we assume that the quasiperiods are incommensurate (i.e.,
that $\tau = \frac{\omega'}{\omega}$ is not a rational number),
the auxiliary variables $Y$ and $\widetilde{Y}$ are multiplicatively incommensurate for generic $y$ (i.e., if $\widetilde{Y}^k Y^l = 1$ for
some integer $k$ and $l$, then $k = l =0$).
Using the f\/inite-dif\/ference equations \eqref{FunEq} our generating function can be
rewritten as a f\/inite product
\begin{gather}
D_{n\omega+m\omega'}(x-y) =
\prod\limits^{n-1}_{k=0}
\big( \widetilde{Y}^{-1} \widetilde{X}
 \tilde{q}^{\frac{n-1}{2}-k}
+ \widetilde{Y} \widetilde{X}^{-1}
(-1)^m \tilde{q}^{-\frac{n-1}{2}+k} \big)
 \notag \\
 \hphantom{D_{n\omega+m\omega'}(x-y)=}{} \times
\prod\limits^{m-1}_{l=0}
\big( Y^{-1} X q^{\frac{m-1}{2}-l} + Y X^{-1}
 (-1)^n q^{-\frac{m-1}{2}+l} \big),
\label{modGenFun}
\end{gather}
where we use the shorthand notation
\begin{gather} \label{Xvar}
X = X(x) = e^{\frac{i \pi}{2\omega}x},\qquad \widetilde{X} = \widetilde{X}(x) = e^{\frac{i \pi}{2\omega'}x}.
\end{gather}
Expanding the Laurent polynomial \eqref{modGenFun} in integer powers of $Y(y)$ and
$\widetilde{Y}(y)$ we extract\linebreak $(n+1)(m+1)$ basis elements of the f\/inite-dimensional
representation given by the mono\-mials
\begin{gather} \label{basisD}
\widetilde{X}^{n-2k} X^{m-2l}
\qquad \text{with}\quad k=0,1,\ldots,n,\quad l=0,1,\ldots,m.
\end{gather}

Let us note that for $s = s_0$ the integral in~\eqref{Wint} diverges.
The divergence is compensated by the normalization factor,
which turns to zero, $A(s_0-\omega'') = 0 $. The ambiguity can be resolved
and for f\/inite-dimensional representations the intertwiner
$D_{-s_0}(\hat p)$ becomes a sum of f\/inite-dif\/ference operators which follows from~\eqref{FunEq}.
One can directly check as well that the basis vectors~\eqref{basisD} are
annihilated by $D_{-s_0}(\hat p)$.

\subsection[An inf\/inite-dimensional ${\rm R}$-operator for the modular double]{An inf\/inite-dimensional $\boldsymbol{{\rm R}}$-operator for the modular double}
\label{3.2}

Now we proceed to integrable structures for the modular double.
The ${\rm L}$-operator is constructed out of
the modular double generators taken in the representation $\pi_s$ (\ref{Gs})~\cite{BT06},
\begin{gather} \label{LBT07}
\mathrm{L}(u\,|\,s) = \left(
\begin{matrix}
e^{\frac{i \pi}{\omega} u} \mathbf{K}_s - e^{-\frac{i \pi}{\omega} u} \mathbf{K}^{-1}_s & (q-q^{-1}) \mathbf{F}_s\vspace{1mm}\\
(q-q^{-1}) \mathbf{E}_s & e^{\frac{i \pi}{\omega} u} \mathbf{K}^{-1}_s - e^{-\frac{i \pi}{\omega} u} \mathbf{K}_s
\end{matrix} \right).
\end{gather}
This ${\rm L}$-operator respects the standard RLL-intertwining relation (cf.~\eqref{RLLYang})
\begin{gather*}
\mathrm{R}_{ab,ef}(u-v) \mathrm{L}_{ec}(u)
\mathrm{L}_{fd}(v) = \mathrm{L}_{bf}(v)
\mathrm{L}_{ae}(u) \mathrm{R}_{ef,cd}(u-v)
\end{gather*}
with $4 \times 4$ trigonometric ${\rm R}$-matrix (cf.~\eqref{q-Yang})
\begin{gather} \label{Rtrig}
\mathrm{R}(u) = \frac{\sin \frac{i \pi}{\omega}(u+\frac{\omega'}{2})}{2\sin \frac{i \pi \omega'}{2\omega}} {\hbox{{1}\kern-.25em\hbox{l}}} \otimes {\hbox{{1}\kern-.25em\hbox{l}}}
+ \tfrac{1}{2} \sigma_1 \otimes \sigma_1 + \tfrac{1}{2} \sigma_2 \otimes \sigma_2
+ \frac{\cos \frac{i \pi}{\omega}(u+\frac{\omega'}{2})}
{2\cos \frac{i \pi \omega'}{2\omega}} \sigma_3 \otimes \sigma_3,
\end{gather}
which is equivalent to the set of commutation relations~(\ref{qsl2}).
The second ${\rm L}$-operator is obtained from
$\mathrm{L}(u)$ by the interchange
$\omega \rightleftarrows \omega'$:
$\widetilde{\mathrm{L}} (u) =
\left. \mathrm{L} (u) \right|_{\omega \rightleftarrows \omega'}$. The same is true for the corresponding $\widetilde{\mathrm{R}}$-matrix.
In the following we indicate formulae only for the ${\rm L}$-operator
(\ref{LBT07}), and all relations for the $\widetilde{\mathrm{L}}$-operator
have the same form with $\omega \rightleftarrows \omega'$.

The ${\rm L}$-operator (\ref{LBT07}) can be represented in the factorized form
\begin{gather} \label{LFact}
\mathrm{L}(u\,|\,s)
= \begin{pmatrix}
U_2 & - U_2^{-1} \\
- U_2^{-1} e^{\frac{i\pi}{\omega}x} & U_2 e^{\frac{i\pi}{\omega}x}
\end{pmatrix}
\begin{pmatrix}
e^{-\frac{i\pi }{2\omega}(p-\omega'')} & 0 \\
0 & e^{\frac{i\pi }{2\omega}(p-\omega'')}
\end{pmatrix}
\begin{pmatrix}
-U_1 & U_1^{-1} e^{-\frac{i\pi}{\omega}x} \\
- U_1^{-1} & U_1 e^{-\frac{i\pi}{\omega}x}
\end{pmatrix},
\\
U_1 = e^{\frac{i\pi}{2 \omega}u_1},\qquad U_2 = e^{\frac{i\pi}{2 \omega}u_2},\nonumber
\end{gather}
where we introduced the ``light-cone'' parameters $u_1$ and $u_2$ instead of
$u$ and $s$
\begin{gather} \label{lambdaparam}
u_1 = u + \tfrac{s}{2} + \tfrac{\omega}{2} - \tfrac{\omega'}{2} , \qquad
u_2 = u - \tfrac{s}{2} + \tfrac{\omega}{2} - \tfrac{\omega'}{2}.
\end{gather}
In the notation $\mathrm{L}(u)$ we omit for simplicity
dependence on the spin parameter~$s$.
Factorization of the ${\rm L}$-operator of the XXZ spin chain
has been introduced in~\cite{BaSt} in relation to the chiral Potts models.
The factorization formula~\eqref{LFact} is completely analogous
to formula~\eqref{Lsl2} for the $\mathrm{SL}(2,\mathbb{C})$ group. The same can be said about
spectral parameters~$u_1$,~$u_2$ in~\eqref{u} and~\eqref{lambdaparam}. However,
although the operators
$\mathrm{L}(u),\widetilde{\mathrm{L}}(u)$ for the modular double look as analogues of~$\mathrm{L}(u)$,~$\bar{\mathrm{L}}(\bar{u})$ for $\mathrm{SL}(2,\mathbb{C})$, in fact,
they are dif\/ferent in their nature.

At the level of ${\rm R}$-operators an analogy with the rational case persists as well.
The general ${\rm R}$-operator acts in the tensor product of two inf\/inite-dimensional
representations $\pi_{s_1}\otimes \pi_{s_2}$~(\ref{Gs}).
It has been found f\/irst in~\cite{BT06}, but the corresponding form of the
${\rm R}$-operator is not suitable for our purposes.
Here we prof\/it from another construction implemented in~\cite{CD14}, where
it has been obtained solving a pair of $\mathrm{R}\mathrm{LL}$-relations
(cf.~\eqref{RLLsl2},~\eqref{RLLsl2'}),
\begin{gather}
 \mathrm{R}_{12}(u-v) \mathrm{L}_1(u\,|\,s_1)
\mathrm{L}_2(v\,|\,s_2) =
\mathrm{L}_1(v\,|\,s_2) \mathrm{L}_2(u\,|\,s_1) \mathrm{R}_{12}(u-v),
\label{RLLMod2} \\
 \mathrm{R}_{12}(u-v) \widetilde{\mathrm{L}}_1(u\,|\,s_1)
\widetilde{\mathrm{L}}_2(v\,|\,s_2) =
\widetilde{\mathrm{L}}_1(v\,|\,s_2) \widetilde{\mathrm{L}}_2(u\,|\,s_1)
\mathrm{R}_{12}(u-v).
\label{RLLMod}
\end{gather}
The spin parameters $s_1$, $s_2$ and the spectral parameters $u$, $v$ appearing
in the $\mathrm{R}\mathrm{LL}$-rela\-tions~\eqref{RLLMod2},~\eqref{RLLMod} are
combined to four ``light-cone'' parameters $u_1$, $u_2$, $v_1$, $v_2$ in accordance with~(\ref{lambdaparam}), i.e.,
\begin{gather*}
u_1 = u + \tfrac{s_1}{2} + \tfrac{\omega}{2} - \tfrac{\omega'}{2} ,\qquad
u_2 = u - \tfrac{s_1}{2} + \tfrac{\omega}{2} - \tfrac{\omega'}{2} ,
\\
v_1 = v + \tfrac{s_2}{2} + \tfrac{\omega}{2} - \tfrac{\omega'}{2} , \qquad
v_2 = v - \tfrac{s_2}{2} + \tfrac{\omega}{2} - \tfrac{\omega'}{2} .
\end{gather*}
The notation $\mathrm{R}_{12}(u-v)$ is a shortened version of
$\mathrm{R}_{12}(u-v\,|\,s_1,s_2)$ or $\mathrm{R}_{12}(u_1,u_2\,|\,v_1,v_2)$
taking into account the spin parameters.

The ${\rm R}$-operator is invariant with respect to the modular double, i.e., it
commutes with the co-product of the generators. More precisely,
\begin{gather} \notag
 [ \mathrm{R}(u), \Delta(\mathbf{K}) ] = [ \mathrm{R}(u), \Delta(\mathbf{E}) ] = [ \mathrm{R}(u), \Delta(\mathbf{F}) ] = 0,
\\ \label{Rtrigcop}
 [ \mathrm{R}(u), \Delta(\widetilde{\mathbf{K}}) ] = [ \mathrm{R}(u), \Delta(\widetilde{\mathbf{E}}) ]
= [ \mathrm{R}(u), \Delta(\widetilde{\mathbf{F}}) ] = 0,
\end{gather}
where we abbreviate the co-product taken in the tensor of representations with spins $s_1$ and $s_2$,
$(\pi_{s_1} \otimes \pi_{s_2}) \circ \Delta$ , to $\Delta$ bearing in mind the
specif\/ied representations.
The co-product is given by the formulae
\begin{gather*}
\Delta(\mathbf{K}) = \mathbf{K}_1 \mathbf{K}_2 ,\qquad
\Delta(\mathbf{E}) = \mathbf{E}_1 \mathbf{K}_2 + \mathbf{K}^{-1}_1 \mathbf{E}_2 ,\qquad
\Delta(\mathbf{F}) = \mathbf{F}_1 \mathbf{K}_2 + \mathbf{K}^{-1}_1 \mathbf{F}_2.
\end{gather*}
Analogous relations take place for $\widetilde{\mathbf{E}}, \widetilde{\mathbf{F}}, \widetilde{\mathbf{K}}$.
The invariance \eqref{Rtrigcop} follows straightforwardly from the $\mathrm{R}\mathrm{L}\mathrm{L}$-relations
\eqref{RLLMod2}, \eqref{RLLMod} subject
to the shift of spectral parameters $u \to u + w$, $v \to v + w$ with arbitrary $w$.

Construction of the general ${\rm R}$-operator for the modular double from~\cite{CD14}
follows the same pattern as for the $\mathrm{SL}(2,\mathbb{C})$ group. It is based on
the elementary intertwining
operators that yield an integral operator representation of the symmetric group $\mathfrak{S}_4$.
Now we repeat with a~slight modif\/ication what has been said in Section~\ref{RSL2} about the
$\mathrm{SL}(2,\mathbb{C})$-invariant ${\rm R}$-operator.
The general ${\rm R}$-operator is a product of four factors which are {\it elementary intertwining} operators
\begin{gather} \label{Rdoubl}
\mathrm{R}_{12}(u-v) = D_{u_2-v_1}(x_{12}) D_{u_1-v_1}(\hat p_2) D_{u_2-v_2}(\hat p_1) D_{u_1-v_2}(x_{12}).
\end{gather}
Here we denote $x_{ij} = x_i - x_j$.
The latter formula has to be compared with \eqref{R} which has the same structure,
only the building blocks are dif\/ferent.
The change $\omega \rightleftarrows \omega'$ does not alter the ${\rm R}$-operator \eqref{Rdoubl}
which satisf\/ies
both $\mathrm{R}\mathrm{L}\mathrm{L}$-relations \eqref{RLLMod2} and \eqref{RLLMod}. Similar to the expres\-sion~\eqref{R},
the representa\-tion~\eqref{Rdoubl} for our inf\/inite-dimensional ${\rm R}$-operator
plays a major role in what follows.
In the next section we f\/ind its reductions to f\/inite-dimensional invariant subspaces.

According to \eqref{Rdoubl} the general ${\rm R}$-operator is a product of four Faddeev--Volkov's ${\rm R}$-mat\-ri\-ces~\cite{VF}.
Applying \eqref{Wint} one can rewrite it explicitly as an integral operator.
Let us note that it is not the only possible form of the ${\rm R}$-operator.
Initially constructed in~\cite{BT06} the ${\rm R}$-operator for the modular double
was obtained in the form which is not convenient enough to address the current problem.
In~\cite{BT06} the ${\rm R}$-operator
appeared in disguise of the $D$-function \eqref{D} and $\operatorname{arcosh}$ of the Casimir operator.
Thus, dealing with such an operator, one has
to decompose tensor products to a sum of irreducible representations
and use the Clebsch--Gordan coef\/f\/icients~\cite{PT}.
The ${\rm R}$-operator in the form~\eqref{Rdoubl} has the virtue of not demanding any
auxiliary information.

In order to justify the chosen terminology for the ${\rm R}$-operator factors in~\eqref{Rtrig},
we indicate here the relations
\begin{gather} \nonumber
D_{u_2-u_1}(\hat p_1) \mathrm{L}_1(u_1,u_2) =
\mathrm{L}_1(u_2,u_1) D_{u_2-u_1}(\hat p_1),
\\ \label{WL2}
D_{v_2-v_1}(\hat p_2) \mathrm{L}_2(v_1,v_2) =
\mathrm{L}_2(v_2,v_1) D_{v_2-v_1}(\hat p_2), \\
D_{u_1-v_2}(x_{12})
\mathrm{L}_1(u_1,u_2)
\mathrm{L}_2(v_1,v_2) =
\mathrm{L}_1(v_2,u_2) \mathrm{L}_2(v_1,u_1)
 D_{u_1-v_2}(x_{12}),
\end{gather}
which have a clear meaning in terms of the permutation group $\mathfrak{S}_4$
and which enable us to check the $\mathrm{R}\mathrm{L}\mathrm{L}$-relations~\eqref{RLLMod2} and~\eqref{RLLMod}.
Similar to~\eqref{rel1} and~\eqref{rel2}, here we assume that the $D$-operators
are acting as $2\times 2$ diagonal matrices proportional to the unit matrix.

The elementary intertwining operators possess a number of peculiar properties.
They satisfy the Coxeter relations (cf.~\eqref{str})
\begin{gather} \label{star-triang}
D_{a} (\hat p_k)  D_{a+b}(x_{12})   D_{b} (\hat p_k) =
D_{b}(x_{12})   D_{a+b}(\hat p_k)   D_{a} (x_{12}),\qquad k =1 ,2 .
\end{gather}
Using these relations one can check~\cite{CD14} that the ${\rm R}$-operator
\eqref{Rdoubl} satisf\/ies the YBE
\begin{gather} \label{YBMod}
\mathrm{R}_{23}(u-v) \mathrm{R}_{12}(u) \mathrm{R}_{23}(v) =
\mathrm{R}_{12}(v) \mathrm{R}_{23}(u) \mathrm{R}_{12}(u-v) .
\end{gather}
Both sides in the latter relation are endomorphisms on the
space $\pi_{s_1} \otimes \pi_{s_2} \otimes \pi_{s_3}$.
For brevity we do not indicate dependence on the spin parameters.

The Coxeter relations~\eqref{star-triang} are equivalent to the star-triangle
relation~\cite{VF} which has three manifestations:
\begin{enumerate}\itemsep=0pt
\item[1)] an integral identity \cite{BT06, BultPhD,Vol05}
\begin{gather*}
A(a) A(b) A(c) \int\limits^{+\infty}_{-\infty} d z D_a(z-z_1) D_b(z-z_2) D_c(z-z_3) \\
{}= D_{-\omega'' - a}(z_2-z_3) D_{-\omega'' - b}(z_3-z_1) D_{- \omega'' - c}(z_1-z_2)
\qquad \mbox{with} \quad a+b+c = -2\omega'' ;
\end{gather*}
\item[2)] a particular point in the image of the operator $D_{-a-\omega''}(\hat p_1)$
(with the same restriction on the parameters as before)
\begin{gather} \label{uniq1doubl}
D_{-a-\omega''}(\hat p_1) \big( D_b(z_{12}) D_c(z_{13})\big) =
\frac{D_{-\omega'' - a}(z_{23}) D_{-\omega'' - b}(z_{13}) D_{- \omega'' - c}(z_{12})}
{A(b)A(c)} ;
\end{gather}
\item[3)] an operator identity
\begin{gather} \label{star-triang-opr}
D_{a} (\hat p) D_{a+b}(x) D_{b} (\hat p) =
D_{b}(x) D_{a+b}(\hat p) D_{a} (x).
\end{gather}
\end{enumerate}

\subsection[Finite-dimensional reduction of the ${\rm R}$-operator]{Finite-dimensional reduction of the $\boldsymbol{{\rm R}}$-operator}
\label{RedMod}

Now we have all ingredients at hand to perform a reduction of the described
${\rm R}$-operator for modular double to a f\/inite-dimensional representation
in one of its tensor factors. The calculation follows precisely the same pattern
as in the $\mathrm{SL}(2,\mathbb{C})$ case (see Section~\ref{RedSL2}). Again the principal
possibility of this reduction is based on the following
relation for the ${\rm R}$-operator~\eqref{Rdoubl}
\begin{gather}
\label{base'}
D_{u_2-u_1}(\hat p_1) \mathbb{R}_{12}(u_1,u_2\,|\,v_1,v_2) = \mathbb{R}_{12}(u_2,u_1\,|\,v_1,v_2) D_{u_2-u_1}(\hat p_1),
\end{gather}
which can be proved using the identity~(\ref{star-triang}).
Here, again, $\mathbb{R}_{12} = \mathrm{P}_{12}\mathrm{R}_{12}$, where
$\mathrm{P}_{12}$ is a~permutation operator.
This relation shows that both, the null-space of the intertwining opera\-tor~$D_{u_2-u_1}(\hat p_1)$ and the image of the intertwining operator
$D_{u_1-u_2}(\hat p_1)$, are mapped onto themselves by our ${\rm R}$-matrix~$\mathbb{R}_{12}$.
Therefore the invariant f\/inite-dimensional subspaces of the null-space
are invariant with respect to the action of the ${\rm R}$-operator as well.

We consider the ${\rm R}$-operator $\mathrm{R}_{12}(u\,|\,s_0,s)$ acting
on the tensor product $\pi_{s_0} \otimes \pi_s$
and introduce the ``light-cone'' parameters (see \eqref{lambdaparam})
\begin{gather}
u_1 = u + \tfrac{s_0}{2} + \tfrac{\omega}{2} - \tfrac{\omega'}{2},\qquad
u_2 = u - \tfrac{s_0}{2} + \tfrac{\omega}{2} - \tfrac{\omega'}{2},\nonumber\\
v_1 = \tfrac{s}{2} + \tfrac{\omega}{2} - \tfrac{\omega'}{2},\qquad
v_2 = - \tfrac{s}{2} + \tfrac{\omega}{2} - \tfrac{\omega'}{2}.\label{lamparam}
\end{gather}
We apply the ${\rm R}$-operator to the function $D_{-\omega''+u_2-u_1}(x_{13})$ in
the f\/irst space. For $s_0 = - \omega''- n \omega - m \omega' $,
$n , m = 0, 1 ,2 , \ldots $, the latter function becomes a generating function
of the f\/inite-dimensional representation in the f\/irst space. However,
for a moment~$s_0$ is assumed to be generic. According to the structure of
${\rm R}$-operator~\eqref{Rdoubl} we consider sequential action of its separate factors.
As the f\/irst step, we apply $D_{u_2-v_2}(\hat p_1) D_{u_1-v_2}(x_{12})$
to $D_{-\omega''+u_2-u_1}(x_{13}) \Phi(x_2)$ and, using formula~\eqref{uniq1doubl}, obtain
\begin{gather}
D_{u_2-v_2}(\hat p_1) D_{u_1-v_2}(x_{12}) D_{-\omega''+u_2-u_1}(x_{13}) \Phi(x_2) \nonumber
\\ \label{modR2expl} \qquad{}
= \frac{A(u_1-u_2)}{A(u_1-v_2)} \cdot D_{u_1-u_2}(x_{12}) D_{-\omega''+v_2-u_1}(x_{13}) D_{u_2-v_2}(x_{23}) \Phi(x_2).
\end{gather}
Further we apply the third factor $D_{u_1-v_1}(\hat p_2)$ of the ${\rm R}$-operator
to both sides of this relation. On the right-hand sides we use the relation
\begin{gather*}
D_{u_1-v_1}(\hat p_2) D_{u_1-u_2}(x_{12}) D_{u_2-v_2}(x_{23}) \Phi(x_2) \nonumber \\
\qquad{} =
A^{-1}(u_1 - v_1) A^{-1}(u_1 - u_2)\cdot D_{u_2-u_1-\omega''}(\hat p_1)
D_{v_1 - u_1 - \omega''}(x_{12}) D_{u_2 - v_2}(x_{13}) \Phi(x_1),
\end{gather*}
which can be easily checked taking into account the integral form
of the intertwiner \eqref{Wint}. In a~full analogy with the $\mathrm{SL}(2,\mathbb{C})$
calculation we traded
a complicated integral operator $D_{u_1-v_1}(\hat p_2)$ for $D_{u_2-u_1-\omega''}(\hat p_1)$
which turns to $D_{n \omega + m \omega'}(\hat p_1)$ in the f\/inite-dimensional setting.
The latter operator is just a sum of the f\/inite-dif\/ference operators which follows
from equations~\eqref{FunEq}.
The substitution $x-y \to \hat p_1$ in~\eqref{modGenFun} yields an explicit expression
for $D_{n \omega + m \omega'}(\hat p_1)$,
\begin{gather}
D_{n\omega+m\omega'}(\hat p) =
\prod\limits^{n-1}_{k=0}
\left( e^{\frac{i \pi}{2\omega'}\hat p}
 \tilde{q}^{\frac{n-1}{2}-k}
+ e^{-\frac{i \pi}{2\omega'}\hat p}
(-1)^m \tilde{q}^{-\frac{n-1}{2}+k} \right)
 \notag \\
\hphantom{D_{n\omega+m\omega'}(\hat p) =}{}
 \times
\prod\limits^{m-1}_{l=0}
\left( e^{\frac{i \pi}{2\omega} \hat p} q^{\frac{m-1}{2}-l} + e^{-\frac{i \pi}{2\omega} \hat p}
 (-1)^n q^{-\frac{m-1}{2}+l} \right).
\end{gather}
The fourth factor of the ${\rm R}$-operator
is inert being the multiplication by a function operator.

Thus the integral ${\rm R}$-operator for the modular double~\eqref{Rdoubl}
acting on the tensor product of two inf\/inite-dimensional representations
$\pi_{s_0} \otimes \pi_{s}$ can be reduced to a f\/inite-dimensional representation
in the f\/irst space for $s_0 = - \omega''- n \omega - m \omega' $, $n, m \in \mathbb{Z}_{\geq 0} $.
It acts on the generating function of f\/inite-dimensional
representation~\eqref{modGenFun} according to the following explicit formula
\begin{gather} \label{redmod}
\mathbb{R}_{12}(u\,|\,s_0,s) \cdot D_{n \omega + m \omega'}(x_{13}) \Phi(x_2) =
c
\cdot D_{u-\frac{s_0}{2}-\frac{s}{2}}(x_{12})
\\ \qquad{}\times
 D_{-u-\frac{s_0}{2}-\frac{s}{2}-\omega''}(x_{23})\cdot
D_{n \omega + m \omega'}(\hat p_{2})\cdot D_{-u-\frac{s_0}{2}+\frac{s}{2}-\omega''}(x_{12})
D_{u-\frac{s_0}{2}+\frac{s}{2}}(x_{23}) \Phi(x_2),
\nonumber
\end{gather}
where the normalization factor is
\begin{gather*}
c^{-1}= A\big(u+ \tfrac{s_0}{2}+\tfrac{s}{2}\big) A\big(u+\tfrac{s_0}{2}-\tfrac{s}{2}\big)
\end{gather*}
and $x_3$ is an auxiliary parameter.

Both sides of the equality \eqref{redmod} can be expanded in integer powers of
the variables~$X_3(x_3) $, $\widetilde{X}_3(x_3)$ (see~\eqref{Xvar}).
This yields simultaneously an expansion in integer powers of the variables
$X_1(x_1) $, $\widetilde{X}_1(x_1)$ (see~\eqref{basisD}), which
form a basis of the f\/inite-dimensional representation of interest.
The resulting formula~\eqref{redmod} is very helpful in applications.
We use it as follows. Firstly, f\/inite-dif\/ference operators in the sum
 $D_{n \omega + m \omega'}(\hat p_2)$ act from the left on the $D$-functions
and shift their arguments.
After these shifts we trade all $D$-functions~\eqref{D} in~\eqref{redmod} for
quantum dilogarithms~\eqref{gamma} and apply the f\/inite-dif\/ference equations~\eqref{gFunEq}. In this way we completely get rid of\/f the quantum dilogarithms.
The f\/inal result contains only trigonometric functions, i.e., a linear
combination of the products of
$X_1(x_1) $, $\widetilde{X}_1(x_1) $, $X_3(x_3) $, $\widetilde{X}_3(x_3) $,
and $\Phi(x_2)$ with the shifted argument. Thus the restriction of the general
${\rm R}$-operator can be represented as an $(n+1)(m+1)$-dimensional
matrix whose entries are f\/inite-dif\/ference operators with the trigonometric coef\/f\/icients.

Formula~\eqref{redmod} constitutes one of the main results of this paper.
It gives a new rich class of solutions of the YBE which are endomorphisms
on a tensor product of f\/inite-dimensional and inf\/inite-dimensional representations
of the modular double specif\/ied in~\eqref{Gs}.

In~\cite{Mangazeev:2014gwa,Mangazeev:2014bqa} an explicit hypergeometric formula
for the ${\rm R}$-matrix of $U_q({\mathfrak{sl}}_2)$ acting on a tensor product of two highest-weight
representations has been presented. It would be interesting to relate our formula
\eqref{redmod} to ${\rm R}$-matrices from these papers. In \cite{KhT} group
theoretical origins of similar factorization formulae were elucidated from the
representation theory of $U_q(\hat{\mathfrak{sl}}_2)$ algebra.

In order to demonstrate how formula \eqref{redmod} works in practice
we recover the ${\rm L}$-operator \eqref{LBT07} out of the
${\rm R}$-operator \eqref{Rdoubl}. With this task in mind,
we choose the spin $s_0 = -\omega'-\omega''$, i.e., f\/ix $n=0$, $m=1$.
The generating function (\ref{modGenFun}) of the $2$-dimensional
representation in the f\/irst space is
\begin{gather*}
D_{\omega'}(x_{13}) = e^{\frac{i \pi}{2 \omega} x_{13}} + e^{-\frac{i \pi}{2 \omega} x_{13}}.
\end{gather*}
Consequently
$\mathbf{e}_1 = e^{\frac{i \pi}{2 \omega} x_{1}}$, $\mathbf{e}_2 = e^{-\frac{i \pi}{2 \omega} x_{1}}$
form a basis of $\mathbb{C}^2$. The f\/inite-dif\/ference operator in~(\ref{redmod}) is
$D_{\omega'}(\hat p_{2}) = e^{\frac{i \pi}{2 \omega} \hat p_{2}} +
e^{-\frac{i \pi}{2 \omega} \hat p_{2}} $.
Up to a normalization factor the right-hand side of~(\ref{redmod}) takes the form
\begin{gather*}
\cosh \tfrac{i \pi}{2 \omega}\left( x_{12} -u + \tfrac{s}{2} - \tfrac{\omega}{2}\right) \cdot
\cosh \tfrac{i \pi}{2 \omega}\left( x_{23} -u - \tfrac{s}{2} - \tfrac{\omega}{2}-\omega'\right)
\cdot \Phi(x_2 - \omega') \\
\qquad{} +
\cosh \tfrac{i \pi}{2 \omega}\left( x_{12} + u - \tfrac{s}{2} + \tfrac{\omega}{2}\right) \cdot
\cosh \tfrac{i \pi}{2 \omega}\left( x_{23} + u + \tfrac{s}{2} + \tfrac{\omega}{2} + \omega'\right)
\cdot \Phi(x_2 + \omega')
\end{gather*}
Expanding this function in terms of
$X^{\pm 1}_1(x_1) X^{\pm 1}_3(x_3) = e^{\frac{i \pi}{2 \omega} (\pm x_{1} \pm x_{3})}$ we obtain
\begin{gather}
\mathbb{R}_{12}\big(u - \omega - \tfrac{\omega'}{2}\big) \mathbf{e}_1 =
c \cdot \big[ \mathbf{e}_1 \big( e^{\frac{i \pi}{\omega} u} \mathbf{K}_s - e^{-\frac{i \pi}{\omega} u} \mathbf{K}^{-1}_s \big)
+ \mathbf{e}_2 \big(q-q^{-1}\big) \mathbf{E}_s \big] ,
\notag \\ \notag
\mathbb{R}_{12}\big(u - \omega - \tfrac{\omega'}{2}\big) \mathbf{e}_2 =
c \cdot \big[ \mathbf{e}_1 \big(q-q^{-1}\big)\mathbf{F}_s +
\mathbf{e}_2 \big( e^{\frac{i \pi}{\omega} u} \mathbf{K}^{-1}_s - e^{-\frac{i \pi}{\omega} u} \mathbf{K}_s \big) \big] .
\end{gather}
Thus we have reproduced the desired result~\eqref{LBT07}. In a similar way one reproduces the $\widetilde{\mathrm{L}}$-operator
at $s_0 = -\omega -\omega''$.
Implementing these reduced ${\rm R}$-operators in the YBE~\eqref{YBMod}
one recovers the $\mathrm{RLL}$-relations~\eqref{RLLMod}.
An explicit matrix factorization formula for $\mathbb{R}_{12}(u\,|\,s_0,s)$
generalizing the ${\rm L}$-operator factorization~\eqref{LFact}
was derived in the followup paper~\cite{CD15}.

A reduction of the ${\rm R}$-operator to f\/inite-dimensional representations in both spaces
can be constructed as well. One just should choose an appropriate discrete value
of the spin~$s$ in~\eqref{redmod} and substitute $\Phi(x_2)$ for the corresponding
generating function. In this way one generates a~number of
f\/inite-dimensional solutions of the YBE including the trigonometric
${\rm R}$-mat\-rix~\eqref{Rtrig} among them.
More precisely, let us f\/ix the spin parameters as
$s_1 = - \omega''- n_1 \omega - m_1 \omega' $, $n_1, m_1 \in \mathbb{Z}_{\geq 0} $,
and $s_2 = - \omega''- n_2 \omega - m_2 \omega' $, $n_2, m_2 \in \mathbb{Z}_{\geq 0} $,
in the f\/irst and second spaces, respectively. Then
\begin{gather}
\mathbb{R}_{12}(u\,|\,s_1,s_2) \cdot D_{n_1 \omega + m_1 \omega'}(x_{13})
D_{n_2 \omega + m_2 \omega'}(x_{24}) =c
\cdot D_{u-\frac{s_1}{2}-\frac{s_2}{2}}(x_{12})
D_{-u-\frac{s_1}{2}-\frac{s_2}{2}-\omega''}(x_{23}) \nonumber
\\ \qquad{}\times
D_{n_1 \omega + m_1 \omega'}(\hat p_{2})\cdot D_{-u-\frac{s_1}{2}+\frac{s_2}{2}
-\omega''}(x_{12})
D_{u-\frac{s_1}{2}+\frac{s_2}{2}}(x_{23}) D_{n_2 \omega + m_2 \omega'}(x_{24})
\label{Rfindim}\end{gather}
is a concise expression for the f\/inite-dimensional (in both spaces) ${\rm R}$-matrix.
After expansion with respect to auxiliary parameters $X_3(x_3) $,
$\widetilde{X}_3(x_3) $, $X_4(x_4) $, and $\widetilde{X}_4(x_4) $
it can be rewritten explicitly
is the form of an $(n_1+1)(m_1+1)(n_2+1)(m_2+1)$-dimensional matrix.

The integral ${\rm R}$-operator exists as well for the $\mathcal{U}_q({\mathfrak{sl}}_2)$-algebra
\cite{Derkachov:2007gr}, which is a~``one-half'' of the modular double.
A reduction of this ${\rm R}$-operator leads to the trigonometric
${\rm L}$-operator as was shown in~\cite{CDKK12}. Derivation of the
corresponding higher-spin f\/inite-dimensional solutions of YBE using
the described reduction procedure will be presented elsewhere.

\subsection[The fusion and symbols for $\mathcal{U}_q({\mathfrak{sl}}_2)$ algebra]{The fusion and symbols for $\boldsymbol{\mathcal{U}_q({\mathfrak{sl}}_2)}$ algebra}
\label{3.4}

Now we would like to show that the reduction result of the previous section
can be derived with the help of the fusion procedure.
In the present section we develop the fusion for the quantum algebra~$\mathcal{U}_q({\mathfrak{sl}}_2)$ and in the next one we consider the modular double.
Our approach is not that well known since we extensively use the symbols of operators.

Similar to the discussion in Section~\ref{FusSL2}
we construct the Lax operator with a f\/inite-dimen\-sio\-nal local quantum space
out of the $q$-deformed Yang's ${\rm R}$-matrix
(remind that the deformation parameter and quasiperiods are related as
$q = e^{i \pi \omega/\omega'}$).
The latter acts on the tensor product of two fundamental representations
and is given by the matrix
\begin{gather}
\mathrm{R}(u) =
\frac{1}{2}
\frac{q^{u+\frac{1}{2}} - q^{-u-\frac{1}{2}}}
{q^{\frac{1}{2}} - q^{-\frac{1}{2}}} {\hbox{{1}\kern-.25em\hbox{l}}} +
\frac{1}{2}\sigma_1\otimes\sigma_1+
\frac{1}{2}\sigma_2\otimes\sigma_2+
\frac{1}{2}\frac{q^{u+\frac{1}{2}} + q^{-u-\frac{1}{2}}}
{q^{\frac{1}{2}} + q^{-\frac{1}{2}}} \sigma_3\otimes\sigma_3
\nonumber\\
\hphantom{\mathrm{R}(u)}{}
= \left(
\begin{matrix}
 \left[u+\frac{1}{2} +
\frac{1}{2} \sigma_3\right]_q & \sigma_{-} \\
 \sigma_{+} & \left[u+\frac{1}{2} -
\frac{1}{2} \sigma_3\right]_q
\end{matrix}
\right),\label{q-Yang}
\end{gather}
where $[x]_q$ is the $q$-number $[x]_q = \frac{q^x-q^{-x}}{q-q^{-1}}$.
Here $\sigma_{\pm}=(\sigma_1\pm i\sigma_2)/2$ and the $q$-number of the matrix~$\sigma_3$
is def\/ined in an evident way, since it is diagonal.

\looseness=-1
In a full analogy with the non-deformed case,
the recipe of \cite{KRS81,KS81} suggests to form an inhomogeneous monodromy matrix out of
the $q$-deformed Yang's ${\rm R}$-matrices and to symmetrize it,
\begin{gather} \label{Rqfus}
\mathrm{R}_{(i_1\ldots i_n)}^{(j_1 \ldots j_n)}(u): = \operatorname{Sym} \mathrm{R}_{i_1}^{j_1}(u) \mathrm{R}_{i_1}^{j_1}(u-1)
 \cdots \mathrm{R}_{i_n}^{j_n}(u-n+1),
\end{gather}
where $\operatorname{Sym}$ implies symmetrization with respect to indices
$(i_1\ldots i_n)$ and $(j_1 \ldots j_n)$
and all these indices refer to the f\/irst space of the ${\rm R}$-matrix~\eqref{q-Yang}
\begin{gather*}
\mathrm{R}_i^j(u) =
\frac{1}{2}
\frac{q^{u+\frac{1}{2}} - q^{-u-\frac{1}{2}}}
{q^{\frac{1}{2}} - q^{-\frac{1}{2}}} \delta_{i}^{j} +
\frac{1}{2}(\sigma_1)^j_{i}\otimes\sigma_1+
\frac{1}{2}(\sigma_2)^j_{i}\otimes\sigma_2+
\frac{1}{2}\frac{q^{u+\frac{1}{2}} + q^{-u-\frac{1}{2}}}
{q^{\frac{1}{2}} + q^{-\frac{1}{2}}} (\sigma_3)^j_{i}\otimes\sigma_3 .
\end{gather*}
The standard way of treating~\eqref{Rqfus} implies construction of the symmetrizer,
i.e., a projector to the highest spin representation in the decomposition of
the product of $n$ fundamental representations. We implement the projection
by means of the auxiliary spinors $\lambda $, $\mu$ that is equivalent to
dealing with the symbols of ${\rm R}$-matrices.
The symbol of \eqref{Rfusfact} (with respect to the local quantum space, not the auxiliary
one) factorizes (see~\eqref{Rfusfact})
\begin{gather}
\mathrm{R}(u\,|\,\lambda,\mu) = \lambda_{i_1} \cdots \lambda_{i_n}
\mathrm{R}_{i_1\ldots i_n}^{j_1 \ldots j_n}(u)
 \mu_{j_1} \cdots \mu_{j_n} \nonumber\\
 \hphantom{\mathrm{R}(u\,|\,\lambda,\mu)}{}
 = \langle\lambda\,|\,\mathrm{R}(u)\,|\,\mu\rangle
\langle\lambda\,|\,\mathrm{R}(u-1)\,|\,\mu\rangle\cdots
\langle\lambda\,|\,\mathrm{R}(u-n+1)\,|\,\mu\rangle\label{YangTrigString}
\end{gather}
to a product of the symbols for $q$-Yang's ${\rm R}$-matrices
$\langle\lambda\,|\,\mathrm{R}(u)\,|\,\mu\rangle = \lambda_{i} \mathrm{R}_{i}^{j}(u) \mu_{j} $,
\begin{gather} \label{RqSymb}
\langle\lambda\,|\,\mathrm{R}(u)\,|\,\mu\rangle =
\left(%
\begin{matrix}
 [u+1]_q \lambda_1\mu_1+[u]_q \lambda_2\mu_2 & \lambda_2\mu_1 \\
 \lambda_1\mu_2 & [u]_q \lambda_1\mu_1+ [u+1]_q \lambda_2\mu_2
\end{matrix}
\right).
\end{gather}
The product of $n$ such matrices is given by
\begin{gather}\label{q-matrix}
\mathrm{R}(u\,|\,\lambda,\mu) = [u]_q[u-1]_q\cdots[u-n+1]_q
\left(
\begin{matrix}
 \mathrm{A}\left(u\,|\,\lambda_1\mu_1,\lambda_2\mu_2\right) & \!\!\mathrm{B}\left(\lambda_1\mu_1,\lambda_2\mu_2\right) \lambda_2\mu_1 \\
 \mathrm{B}\left(\lambda_2\mu_2,\lambda_1\mu_1\right) \lambda_1\mu_2\!\! &
 \mathrm{A}\left(u\,|\,\lambda_2\mu_2,\lambda_1\mu_1\right)
\end{matrix}
\right),\!\!\!
\end{gather}
where the functions $A$ and $B$ have the following form
\begin{gather}
\mathrm{A}\left(u\,|\,\lambda_1\mu_1,\lambda_2\mu_2\right) =
\sum_{k=0}^n \frac{n!}{k!(n-k)!} [u+1+k-n]_q
(\lambda_1\mu_1)^{n-k} (\lambda_2\mu_2)^{k}
\nonumber\\
\hphantom{\mathrm{A}\left(u\,|\,\lambda_1\mu_1,\lambda_2\mu_2\right)}{}
=\big[u+1-\tfrac{n}{2}+\tfrac{1}{2}(\lambda_1\partial_{\lambda_1}-
\lambda_2\partial_{\lambda_2})\big]_q
(\lambda_1\mu_1+\lambda_2\mu_2)^n,\label{Afun}
\\ \nonumber
\mathrm{B}\left(\lambda_1\mu_1,\lambda_2\mu_2\right) =
\sum_{k=0}^n \frac{n!}{k!(n-k)!} [k]_q
(\lambda_1\mu_1)^{k-1} (\lambda_2\mu_2)^{n-k} =
\\ \label{B}
\hphantom{\mathrm{B}\left(\lambda_1\mu_1,\lambda_2\mu_2\right)}{}
= \frac{1}{\lambda_1\mu_1}\left[\lambda_1\partial_{\lambda_1}\right]_q
(\lambda_1\mu_1+\lambda_2\mu_2)^n.
\end{gather}
The summation formulae in~\eqref{Afun}, \eqref{B} facilitate
reconstruction of operators from the symbolic entries of the matrix~\eqref{q-matrix}.
In analogy with the non-deformed case we again remove the inessential normalization
factor $r_n(u)=[u]_q[u-1]_q\cdots[u-n+1]_q$ and shift the
spectral parameter $u \to u-1+\frac{n}{2}$ to obtain a symbol
of the ${\rm L}$-operator
\begin{gather*}
\mathrm{L}(u\,|\,\lambda,\mu) = r^{-1}_n(u) \mathrm{R}\big(u-1+\tfrac{n}{2}\,|\,\lambda,\mu\big)
\\
\hphantom{\mathrm{L}(u\,|\,\lambda,\mu)}{}
=
\left(%
\begin{matrix}
 \left[u+\frac{1}{2}(\lambda_1\partial_{\lambda_1}-
\lambda_2\partial_{\lambda_2})\right]_q & \frac{\lambda_2}{\lambda_1} [\lambda_1\partial_{\lambda_1}]_q\\
 \frac{\lambda_1}{\lambda_2} [\lambda_2\partial_{\lambda_2}]_q &
 \left[u+\frac{1}{2}(\lambda_2\partial_{\lambda_2}-
\lambda_1\partial_{\lambda_1})\right]_q
\end{matrix}
\right) \langle\lambda\,|\,\mu\rangle^n.
\end{gather*}
Since $\langle\lambda\,|\,\mu\rangle^n$ is a symbol of the identity operator,
applying \eqref{TPsispinor}, we immediately recover the familiar Lax operator,
\begin{gather} \label{LJtrig}
\mathrm{L}(u) = \left(
\begin{matrix}
 [u+J_3]_q & J_{-} \\
 J_{+} & [u-J_3]_q
\end{matrix}
\right),
\end{gather}
where the generators of $\mathcal{U}_q({\mathfrak{sl}}_2)$ are realized by the
f\/inite-dif\/ference operators $J_{-}$, $J_{+}$, $q^{\pm J_{3}}$ in two variables,
\begin{gather}
J_{-} = \tfrac{\lambda_2}{\lambda_1} [\lambda_1\partial_{\lambda_1}]_q, \qquad
J_{+} = \tfrac{\lambda_1}{\lambda_2} [\lambda_2\partial_{\lambda_2}]_q, \qquad
J_3 = \tfrac{1}{2}(\lambda_1\partial_{\lambda_1}-
\lambda_2\partial_{\lambda_2}). \label{J-+3}
\end{gather}
One can check that they do respect commutation relations of $\mathcal{U}_q({\mathfrak{sl}}_2)$
\begin{gather*}
J_+ J_- - J_+ J_- = [2 J_3]_q,\qquad J_3 J_\pm - J_\pm J_3 = \pm J_{\pm}.
\end{gather*}

Let us remind that the representation is def\/ined on the space of homogeneous polynomials of
two variables $\lambda_1$, $\lambda_2$ of degree~$n$ (see~\eqref{PsiHom}).
In order to proceed to the space of polynomials of one variable we choose
$\lambda_1 = -z $, $\lambda_2 = 1$, so that the generators~\eqref{J-+3} take
the conventional form
\begin{gather} \label{JgenTrig}
J_{-} = -\tfrac{1}{z} [z\partial]_q, \qquad
J_{+} = z [z\partial-n]_q, \qquad
J_3 = z\partial-\tfrac{n}{2}.
\end{gather}

We close this section by an alternative calculation of the symbol
\eqref{YangTrigString} which follows the pattern at the end of Section~\ref{FusSL2}.
The main merit of the following calculation
is that it can be generalized to the elliptic case~\cite{CDS}.
First of all we f\/ix the auxiliary spinor $\lambda_1 = -z$, $\lambda_2=1$
and use the realization of spin~$\frac{1}{2}$ generators (cf.~\eqref{JgenTrig})
\begin{gather*}
 J_{-} = -\tfrac{1}{z}[z\partial]_q, \qquad J_{+} = z[z\partial - 1]_q,
 \qquad J_{3} = z\partial -\tfrac{1}{2},
\end{gather*}
which act in the two-dimensional space of linear functions $\psi(z) = a_1 z+a_0$.
In the basis $\mathbf{e}_1 = -z$, $\mathbf{e}_2=1$,
the matrices of the generators coincide with the Pauli-matrices
\begin{gather}\label{basis1'}
J_{\pm}\left(\mathbf{e}_1, \mathbf{e}_2\right)=
\left(\mathbf{e}_1, \mathbf{e}_2\right) \sigma_{\pm}, \qquad
J_{3}\left(\mathbf{e}_1, \mathbf{e}_2\right) =
\left(\mathbf{e}_1, \mathbf{e}_2\right) \tfrac{1}{2}\sigma_{3}.
\end{gather}
The fusion procedure enables us to derive the L-operator~\eqref{LJtrig} together with
representation of the spin $\frac{n}{2}$ generators~\eqref{JgenTrig}
acting in the $(n+1)$-dimensional space of polynomials $\psi(z) = a_n z^n+\cdots+a_0$.
Relations~(\ref{basis1'}) are equivalent to
\begin{gather*}
(-z, 1) \sigma_{\pm} = J_{\pm}(-z, 1), \qquad
(-z, 1) \tfrac{1}{2}\sigma_{3} =
J_{3}(-z, 1)
\end{gather*}
that enables us to represent the symbol of $q$-Yang's ${\rm R}$-matrix~\eqref{q-Yang}
as a matrix dif\/ference operator acting on the symbol of the identity operator (cf.~\eqref{RqSymb}),
\begin{gather*}
\langle\lambda\,|\,\mathrm{R}(u)\,|\,\mu\rangle
= \left(
\begin{matrix}
 (-z, 1)
 \left[u+\tfrac{1}{2} +
\tfrac{1}{2} \sigma_3\right]_q\,|\,\mu\rangle &
(-z, 1)\sigma_{-}\,|\,\mu\rangle \\
 (-z, 1)\sigma_{+}\,|\,\mu\rangle & (-z, 1)\left[u+\tfrac{1}{2} -
\tfrac{1}{2} \sigma_3\right]_q\,|\,\mu\rangle
\end{matrix}%
\right)
\\
\hphantom{\langle\lambda\,|\,\mathrm{R}(u)\,|\,\mu\rangle}{}
=
\left(
\begin{matrix}
 [u+z\partial]_q & -\frac{1}{z}[z\partial]_q \\
 z[z\partial-1]_q& [u+1-z\partial]_q
\end{matrix}
\right)\left(\mu_2-\mu_1 z\right).
\end{gather*}
This operator is just the trigonometric L-operator~\eqref{LJtrig} for the
spin $\frac{1}{2}$ representation. The crucial observation is that it can be
factorized respecting a special ordering of~$z$ and~$\partial$,
\begin{gather} 
\langle\lambda\,|\,\mathrm{R}(u)\,|\,\mu\rangle
=
\frac{1}{q-q^{-1}}\begin{pmatrix}
1 & 1 \\
z q^{-u-1} & z q^{u+1}
\end{pmatrix}
\begin{pmatrix}
q^{z_1\partial_1} & 0 \\
0 & q^{-z_1\partial_1}
\end{pmatrix}
\begin{pmatrix}
q^{u} & -z^{-1} \\
- q^{-u} & z^{-1}
\end{pmatrix}\nonumber\\
\hphantom{\langle\lambda\,|\,\mathrm{R}(u)\,|\,\mu\rangle=}{}\times
\left.
\left(\mu_2-\mu_1 z_1\right)\right|_{z_1=z}.\label{LnewFactTrig}
\end{gather}
Formula \eqref{LnewFactTrig} represents a trigonometric deformation of the factorization formula~\eqref{LnewFact}.

The derived formula enables us to simplify the product of
two consecutive symbols from~\eqref{YangTrigString} since a pair of
adjacent matrix factors is cancelled
\begin{gather*}
\langle\lambda\,|\,\mathrm{R}(u)\,|\,\mu\rangle
\langle\lambda\,|\,\mathrm{R}(u-1)\,|\,\mu\rangle = \frac{1}{(q-q^{-1})^2} \begin{pmatrix}
1 & 1 \\
z q^{-u-1} & z q^{u+1}
\end{pmatrix}
\begin{pmatrix}
q^{z_1\partial_1} & 0 \\
0 & q^{-z_1\partial_1}
\end{pmatrix}
\\ \qquad {}\times \begin{pmatrix}
q^{u} & -z^{-1} \\
- q^{-u} & z^{-1}
\end{pmatrix}
\begin{pmatrix}
1 & 1 \\
z q^{-u} & z q^{u}
\end{pmatrix}
\begin{pmatrix}
q^{z_2\partial_2} & 0 \\
0 & q^{-z_2\partial_2}
\end{pmatrix}
\begin{pmatrix}
q^{u-1} & -z^{-1} \\
- q^{-u+1} & z^{-1}
\end{pmatrix}
\\
\left.
\qquad {}\times \left(\mu_2-\mu_1 z_1\right)
\left(\mu_2-\mu_1 z_2\right)
\right|_{z_1=z_2=z}
=
\frac{[u]_q}{q-q^{-1}}\begin{pmatrix}
1 & 1 \\
z q^{-u-1} & z q^{u+1}
\end{pmatrix}
\\ \qquad {}\times
\begin{pmatrix}
q^{z_1\partial_1+z_2\partial_2} & 0 \\
0 & q^{-z_1\partial_1-z_2\partial_2}
\end{pmatrix}
\begin{pmatrix}
q^{u-1} & -z^{-1} \\
- q^{-u+1} & z^{-1}
\end{pmatrix}
\left.
\left(\mu_2-\mu_1 z_1\right)
\left(\mu_2-\mu_1 z_2\right)
\right|_{z_1=z_2=z}.
\end{gather*}
The generalization of this formula is obvious.
Thus the product of symbols \eqref{YangTrigString} is equal to
\begin{gather} \notag
\mathrm{R}(u\,|\,\lambda,\mu)
= r_n(u) \frac{1}{q-q^{-1}}
\begin{pmatrix}
1 & 1 \\
z q^{-u-1} & z q^{u+1}
\end{pmatrix}
\begin{pmatrix}
q^{z_1\partial_1+\cdots+z_n\partial_n} & 0 \\
0 & q^{-z_1\partial_1-\cdots-z_n\partial_n}
\end{pmatrix}
 \\
 \hphantom{\mathrm{R}(u\,|\,\lambda,\mu)=}{} \times
\begin{pmatrix}
q^{u-n+1} & -z^{-1} \\
- q^{-u+n-1} & z^{-1}
\end{pmatrix}\left.
\left(\mu_2-\mu_1 z_1\right)\cdots
\left(\mu_2-\mu_1 z_n\right)
\right|_{z_1=\cdots=z_n=z} . \label{Rproduct}
\end{gather}
Then we need to get rid of\/f the taken special ordering in~\eqref{Rproduct}.
To that end we apply an evident formula
\begin{gather*}
q^{\pm(z_1\partial_1+\cdots+z_n\partial_n)}\left.
\left(\mu_2-\mu_1 z_1\right)\cdots
\left(\mu_2-\mu_1 z_n\right)
\right|_{z_1=\cdots=z_n=z} =
q^{\pm z\partial} \left(\mu_2-\mu_1 z\right)^n,
\end{gather*}
which enables us to cast the matrix product on the right-hand side of \eqref{Rproduct} in the form
\begin{gather*}
r_n(u)
\begin{pmatrix}
[u-n+1 +z_1\partial_1+\cdots+z_n\partial_n]_q & -\frac{1}{z}[z_1\partial_1+\cdots+z_n\partial_n]_q \\
z[z_1\partial_1+\cdots+z_n\partial_n-n]_q & [u+1 -z_1\partial_1-\cdots-z_n\partial_n]_q
\end{pmatrix}
\\
\qquad\quad{}\times \left.
\left(\mu_2-\mu_1 z_1\right)\cdots
\left(\mu_2-\mu_1 z_n\right)
\right|_{z_1=\cdots=z_n=z}
\\
\qquad{}
= r_n(u)
\begin{pmatrix}
[u-n+1 +z\partial]_q& -\frac{1}{z}[z\partial]_q \\
z[z\partial-n]_q & [u+1 -z\partial]_q
\end{pmatrix}
\left(\mu_2-\mu_1 z\right)^n.
\end{gather*}
Thus the f\/inal result for the symbol of the ``fused'' $q$-Yang ${\rm R}$-matrices~\eqref{Rqfus} is
\begin{gather*}
\mathrm{R}(u\,|\,\lambda,\mu)
= r_n(u)
 \begin{pmatrix}
[u+1- \frac{n}{2}+J_3]_q& J_- \\
J_+ & [u+1- \frac{n}{2}-J_3]_q
\end{pmatrix}
\left(\mu_2-\mu_1 z\right)^n ,
\end{gather*}
where the generators $J_{\pm}$, $J_3$ in spin $\frac{n}{2}$
representation are given by the expression~(\ref{JgenTrig}).

\subsection{Fusion construction for the modular double}
\label{fusMD}

The fusion procedure for the modular double closely follows the construction from Section~\ref{fusSL2}.
One forms inhomogeneous monodromy matrix out of the ${\rm L}$-operators and then
symmetrizes it over the spinor indices
resulting in a f\/inite-dimensional (in one of the spaces) higher-spin ${\rm L}$-operator.

Again, instead of working with the higher-rank tensors we introduce auxiliary spinors~$\lambda_i$, $\widetilde{\lambda}_i$,~$\mu_j$, and $\widetilde{\mu}_j$ and contract
them with the monodromy matrix according to~\eqref{Tlam}.
The homoge\-nei\-ty~\eqref{PsiHom} implies that there are redundant variables.
We get rid of\/f them by choosing the gauge $\lambda_1 \lambda_2 = -1$,
$\mu_1 \mu_2 = -1$ that is equivalent to the parametrization of the spinors
by means of the independent variables~$a$ and~$b$ as follows
\begin{gather}
\lambda_1 = \lambda_1(a) = e^{\frac{i \pi}{2\omega}a},\qquad
\lambda_2 = \lambda_2(a) = - e^{-\frac{i \pi}{2\omega}a},\nonumber\\
\mu_1 = \mu_1(b) = e^{\frac{i \pi}{2\omega}b},\qquad
\mu_2 = \mu_2(b) = - e^{-\frac{i \pi}{2\omega}b}.\label{lambmu}
\end{gather}
Analogous relations hold for the spinors $\widetilde{\lambda}$ and $\widetilde{\mu}$
obtained from~\eqref{lambmu} after the interchange $\omega \rightleftarrows \omega'$
with the same~$a$ and~$b$.
Since we assume that the ratio of quasiperiods $\tau$
is not rational, $\lambda$~and~$\widetilde{\lambda}$ are multiplicatively incommensurate
for generic~$a$ and the same is true for~$\mu$ and~$\widetilde{\mu}$ for generic~$b$.

Further, we form symbols of the ${\rm L}$-operators~\eqref{LBT07} (i.e., some
scalar operators) contracting them in the matrix space with the auxiliary
spinors\footnote{We are grateful to D.~Karakhanyan
and R.~Kirschner for a discussion on this point.}
\begin{gather} \label{LamMod}
\lambda_{i} \mathrm{L}^{j}_{i}(u) \mu_{j} = \Lambda(u,\lambda,\mu),\qquad
\widetilde{\lambda}_{i} \widetilde{\mathrm{L}}^{j}_{i}(u) \widetilde{\mu}_{j}
= \widetilde{\Lambda}(u,\widetilde{\lambda},\widetilde{\mu}) .
\end{gather}
Taking into account that
$D_{\omega'}(\hat p) = e^{-\frac{i \pi }{2\omega} \hat p} + e^{\frac{i \pi }{2\omega} \hat p}$
(see \eqref{gFunEq}, \eqref{D}), one can easily check the equality
\begin{gather*}
D_{u_2+\omega'}(x - a ) D_{-u_1}(x + b ) D_{\omega'}(\hat p)
D_{-u_2}(x - a) D_{u_1+\omega'}(x + b ) = i \Lambda(u).
\end{gather*}
It will be helpful to rewrite this formula in a slightly dif\/ferent form by
means of the operator star-triangle relation \eqref{star-triang-opr}
\begin{gather} \label{24}
D_{u_2+\omega'}(x - a) D_{u_1+\omega'}(\hat p) D_{\omega'}(x + b ) D_{-u_1}(\hat p)
D_{-u_2}(x - a) = i \Lambda(u).
\end{gather}
The analogous relation is valid for $\widetilde{\Lambda}(u) $, which is obtained
after the permutation $\omega \rightleftarrows \omega'$.
The derived formula is reminiscent to the ${\rm L}$-operator factorization \eqref{LFact}.
Now we form a string out of the symbols $\Lambda$ and $\widetilde{\Lambda}$ \eqref{24}
with the shifted spectral parameters,
\begin{gather*}
\mathrm{R}_{\text{fus}}(u\,|\,\lambda,\widetilde{\lambda},\mu,\widetilde{\mu}) =
\Lambda(u)\Lambda(u-\omega')\cdots \Lambda(u-(m-1)\omega')
\\ \hphantom{\mathrm{R}_{\text{fus}}(u\,|\,\lambda,\widetilde{\lambda},\mu,\widetilde{\mu}) =}{}\times
\widetilde{\Lambda}(u-(m-1)\omega'-\omega)
\cdots \widetilde{\Lambda}(u-(m-1)\omega'-n\omega).
\end{gather*}
In view of the ref\/lection formula~\eqref{Dev} and relation~\eqref{24} this product
can be recast to the form
\begin{gather}
\mathrm{R}_{\text{fus}}(u\,|\,\lambda,\widetilde{\lambda},\mu,\widetilde{\mu}) =
D_{u_2+\omega'}(x-a)D_{u_1+\omega'}(\hat p)
\left[i D_{\omega'}(x+b)\right]^m \left[i D_{\omega}(x+b)\right]^n
\nonumber\\ \hphantom{\mathrm{R}_{\text{fus}}(u\,|\,\lambda,\widetilde{\lambda},\mu,\widetilde{\mu}) =}{}\times
D_{-u_1+(m-1)\omega'+n\omega}(\hat p)D_{-u_2+(m-1)\omega'+n\omega}(x-a).\label{LamModline}
\end{gather}

Finally, we reconstruct the operator of interest from its symbol
using formula \eqref{TPsispinor}, which results in the representation
\begin{gather} \label{fusMod}
\left[ \mathrm{R}_{\text{fus}}(u) \Phi \right](\lambda,\widetilde{\lambda}\,|\,x)
= \left. \mathrm{R}_{\text{fus}}(u\,|\,\lambda,\widetilde{\lambda},\partial_{\mu},\partial_{\widetilde{\mu}})
\Phi(\mu,\widetilde{\mu}\,|\,x) \right|_{\mu = \widetilde{\mu}= 0},
\end{gather}
where the symbol $\mathrm{R}_{\text{fus}}$ is f\/ixed in \eqref{LamModline}.
Let us stress once more that the fusion formula \eqref{fusMod} is completely analogous
to the $\mathrm{SL}(2,\mathbb{C})$ group case~\eqref{fussl2}.
The higher-spin ${\rm R}$-operator acts on a function
$\Phi(\lambda,\widetilde{\lambda}\,|\,x)$ having the homogeneity degrees~$m$ in $\lambda$ and~$n$ in $\widetilde{\lambda}$, respectively.
In \eqref{fusMod} one has dif\/ferentiations over spinors~$\mu $,~$\widetilde{\mu} $, but
the operator $\mathrm{R}_{\text{fus}}$ \eqref{LamModline} formally depends on~$b$
and not on the exponential of~$b $.
In order to see that there is no contradiction, we note
that according to the def\/initions~\eqref{LamMod} $\Lambda$ and $\widetilde{\Lambda}$ are
linear in spinors. Consequently $\mathrm{R}_{\text{fus}}$~\eqref{LamModline},
being a product of them, has to be polynomial in spinors. This can be checked
directly as well. Recalling the def\/inition of~$\mu $, $\widetilde{\mu} $
\eqref{lambmu} and
\begin{gather} \label{DD}
D_{\omega'}(x+b) = \mu_1 e^{\frac{i \pi}{2 \omega}x}
- \mu_2 e^{-\frac{i \pi}{2 \omega}x} , \qquad
D_{\omega}(x+b) = \widetilde{\mu}_1 e^{\frac{i \pi}
{2 \omega'}x} - \widetilde{\mu}_2 e^{-\frac{i \pi}{2 \omega'}x},
\end{gather}
we conclude that $\mathrm{R}_{\text{fus}}$ in \eqref{LamModline} depends
polynomially on $\mu$ and $\widetilde{\mu}$.
Thus the fusion formu\-lae~\eqref{LamModline} and~\eqref{fusMod} match to each other.

The right-hand side of \eqref{LamModline} explicitly depends on~$a $, so its polynomiality
in spinors $\lambda $, $\widetilde{\lambda}$ is not obvious at all.
It is necessary to demonstrate it explicitly.
Furthermore, we need to compa\-re~\eqref{fusMod} with the reduction formula \eqref{redmod},
since both give rise to a higher-spin ${\rm R}$-operator.
We will accomplish both tasks if we show that the ${\rm R}$-operators do coincide.
Thus we take the generating function
$D_{n \omega + m\omega'}(a-y)$ of a f\/inite-dimensional representation and
act upon it by the ``fused'' ${\rm R}$-operator in the f\/irst space according
to the prescription~\eqref{fusMod}.
The generating function with the auxiliary parameter $y$
explicitly depends on $\lambda$ and $\widetilde{\lambda}$ (see~\eqref{modGenFun}),
\begin{gather}
D_{n\omega+m\omega'}(a-y) =
\prod\limits^{n-1}_{k=0}
\left( \widetilde{\lambda}_1 e^{-\frac{i \pi}{2\omega'}y}
 \tilde{q}^{\frac{n-1}{2}-k}
- (-1)^m \widetilde{\lambda}_2
e^{\frac{i \pi}{2\omega'}y}
 \tilde{q}^{-\frac{n-1}{2}+k} \right)
\cdot \notag \\
\hphantom{D_{n\omega+m\omega'}(a-y) =}{}\times
\prod\limits^{m-1}_{p=0}
\left( \lambda_1 e^{-\frac{i \pi}{2\omega}y}
q^{\frac{m-1}{2}-p}
- (-1)^n \lambda_2 e^{\frac{i \pi}{2\omega}y}
q^{-\frac{m-1}{2}+p} \right),\label{genf}
\end{gather}
and has the homogeneity degrees $m$ in $\lambda$ and $n$ in $\widetilde{\lambda}$,
respectively. Now, using the relations (see~\eqref{DD},~\eqref{genf})
\begin{gather}
\left.\left[ D_{\omega'}(x+b) \right]^m \left[ D_{\omega}(x+b) \right]^n
\right|_{\mu\to\partial_{\mu},\widetilde{\mu}\to\partial_{\widetilde{\mu}}}
D_{n \omega + m\omega'}(b-y)
= n!m! D_{n\omega + m\omega'}(x-y),
\end{gather}
we can perform dif\/ferentiations over spinors in \eqref{fusMod}
that immediately yield the desired result
\begin{gather}
\mathrm{R}_{\text{fus}}\big(u+{\tfrac{n\omega}{2}} + {\tfrac{m\omega'}{2}}\big) D_{n \omega + m\omega'}(a-y) =
i^{n+m}n!m! D_{u-\frac{s_0}{2}-\frac{s}{2}}(x-a)
D_{u-\frac{s_0}{2}+\frac{s}{2}}(\hat p) \notag \\
\quad{}\times D_{n\omega + m\omega'}(x-y)
D_{-u - \frac{s_0}{2} - \frac{s}{2} - \omega''}(\hat p)
D_{-u - \frac{s_0}{2} + \frac{s}{2} - \omega''}(x-a)
= i^{n+m}n!m!
D_{u-\frac{s_0}{2}-\frac{s}{2}}(x-a) \notag \\
\quad{}\times
 D_{-u - \frac{s_0}{2} - \frac{s}{2} - \omega''}(x-y)
D_{n\omega + m\omega'}(\hat p)
D_{u-\frac{s_0}{2}+\frac{s}{2}}(x-y)
D_{-u - \frac{s_0}{2} + \frac{s}{2} - \omega''}(x-a),
\label{fusgenmod}
\end{gather}
where at the last step we prof\/ited from the operator star-triangle relation~\eqref{star-triang-opr}.
Identifying the variables $a = x_1$, $x = x_2$, $y = x_3$,
we f\/ind a nice agreement of the fusion formula~\eqref{fusgenmod} with the
reduction formula~\eqref{redmod}.
Thus both approaches are equivalent and yield identical results.

\subsection*{Acknowledgement}

We thank the referees for useful remarks to the paper.
This work is supported by the Russian Science Foundation
(project no.~14-11-00598).

\LastPageEnding

\end{document}